%% file: main.tex
\documentclass[ALICE,manyauthors]{cernphprep}

\newcommand{\pp}{p\kern-0.05em p}
\newcommand{\ppbar}{\mathrm{p}\kern-0.05em \overline{\mathrm{p}}}
\newcommand{\pPb}{\ensuremath{\mbox{p--Pb}}}
\newcommand{\PbPb}{\ensuremath{\mbox{Pb--Pb}}}
\newcommand{\GeVc}{\ensuremath{\mathrm{GeV}\kern-0.05em/\kern-0.02em c}}

\newcommand{\pT}{\ensuremath{p_{\mathrm{T}}}}
\newcommand{\pTi}{\ensuremath{p_{\mathrm{T},i}}}
\newcommand{\pTtrack}{\ensuremath{p_{\mathrm{T,track}}}}

\newcommand{\etajet}{\ensuremath{\eta_{\mathrm{jet}}}}
\newcommand{\pTjet}{\ensuremath{p_{\mathrm{T}}^{\mathrm{jet}}}}
\newcommand{\pTchjet}{\ensuremath{p_{\mathrm{T}}^{\mathrm{ch\; jet}}}}
\newcommand{\pTtruth}{\ensuremath{p_{\mathrm{T,truth}}^{\mathrm{ch\; jet}}}}
\newcommand{\pTdet}{\ensuremath{p_{\mathrm{T,det}}^{\mathrm{ch\; jet}}}}

\newcommand{\ang}{\ensuremath{\lambda_{\alpha}}}
\newcommand{\angtruth}{\ensuremath{\lambda_{\alpha,\mathrm{truth}}}}
\newcommand{\angdet}{\ensuremath{\lambda_{\alpha,\mathrm{det}}}}
\newcommand{\angNP}{\ensuremath{\lambda_{\alpha}^{\mathrm{NP}}}}

\newcommand{\angsd}{\ensuremath{\lambda_{\alpha,\mathrm{g}}}}
\newcommand{\angsdNP}{\ensuremath{\lambda_{\alpha,\mathrm{g}}^{\mathrm{NP}}}}

\newcommand{\zcut}{\ensuremath{z_{\mathrm{cut}}}}
\newcommand{\betasd}{\ensuremath{\beta}}
\newcommand{\pTsub}{\ensuremath{p_{\mathrm{T,subleading}}}}
\newcommand{\pTlead}{\ensuremath{p_{\mathrm{T,leading}}}}

\usepackage[comma,square,numbers,sort&compress]{natbib}
\usepackage{hyperref}
\usepackage{lineno}
\usepackage{amsmath}
\usepackage{bigints}  % larger integral signs
\usepackage{afterpage}  % big figures can have their own page
\usepackage[T1]{fontenc}

\begin{document}%

%%%%%%%%%%%%%%%  Title page %%%%%%%%%%%%%%%%%%%%%%%%
\begin{titlepage}
\PHyear{2021}
\PHnumber{145}      % required, will be obtained from PH
\PHdate{18 July}  % required, will be obtained from PH
%

%%% Put your own title + short title here:
\title{Measurements of the groomed and ungroomed jet angularities \mbox{in \pp{} collisions at $\mathbf{\sqrt{\textit{s}}} \mathbf{\;= 5.02}$ TeV}}

\ShortTitle{Measurements of the jet angularities in \pp{} collisions at $\sqrt{s} = 5.02$ TeV}   % appears on right page headers

%%% Do not change the next lines
\Collaboration{ALICE Collaboration\thanks{See Appendix~\ref{app:collab} for the list of collaboration members}}
\ShortAuthor{ALICE Collaboration} % appears on left page headers, do not change

\begin{abstract}
The jet angularities are a class of jet substructure observables which characterize the angular and momentum distribution of particles within jets. 
These observables are sensitive to momentum scales ranging from perturbative hard scatterings to nonperturbative fragmentation into final-state hadrons.
We report measurements of several groomed and ungroomed
%infrared- and collinear-safe
jet angularities in pp collisions at $\sqrt{s} = 5.02$ TeV with the ALICE detector. 
Jets are reconstructed using charged particle tracks at midrapidity
($|\eta| < 0.9$).
The anti-$k_{\rm T}$ algorithm is used with jet resolution parameters $R = 0.2$ and $R = 0.4$ for several transverse momentum $p_{\rm T}^{\text{ch jet}}$ intervals in the 20$-$100 GeV/$c$ range.
Using the jet grooming algorithm Soft Drop, the sensitivity to softer, wide-angle processes, as well as the underlying event, can be reduced in a way which is well-controlled in theoretical calculations.
We report the ungroomed jet angularities, $\lambda_{\alpha}$, and groomed jet angularities, $\lambda_{\alpha\text{,g}}$, to investigate the interplay between perturbative and nonperturbative effects at low jet momenta. Various angular exponent parameters $\alpha = 1$, 1.5, 2, and 3 are used to systematically vary the sensitivity of the observable to collinear and soft radiation. 
Results are compared to analytical predictions at next-to-leading-logarithmic accuracy, which provide a generally good description of the data in the perturbative regime but exhibit discrepancies in the nonperturbative regime. 
Moreover, these measurements serve as a baseline for future ones in heavy-ion collisions by providing new insight into the interplay between perturbative and nonperturbative effects in the angular and momentum substructure of jets. 
They supply crucial guidance on the selection of jet resolution parameter, jet transverse momentum, and angular scaling variable for jet quenching studies.

\end{abstract}
\end{titlepage}
\setcounter{page}{2}

\input{Ch1_Introduction} 
\input{Ch2_ExperimentalSetup}
\input{Ch3_DataAnalysis} 
\input{Ch4_SystematicUncertainties}

\input{Ch5_Results} 
\input{Ch6_Conclusion} 
%\clearpage
%
%

%%%%% acknowledgements
\newenvironment{acknowledgement}{\relax}{\relax}
\begin{acknowledgement}
\section*{Acknowledgements}
We gratefully acknowledge Kyle Lee and Felix Ringer for providing theoretical predictions,
and for valuable discussions regarding the comparison of these predictions to our measurements.
\input{fa_2021-07-05.tex}    %%%%%%% done by webmaster team
\end{acknowledgement}

%%%%%%%% Bibliography (In case of using bibtex generate the bbl requested by arXiv)
%\bibliographystyle{elsarticle-num}
\bibliographystyle{utphys}   % Remember we use title in the biblio
\bibliography{main}
%\input {bibliography.tex}  

%%%%%%%%% appendix with author list
\newpage
\appendix
\input{AppendixA}               %%%%%%%%%%% put your appendices here
\section{The ALICE Collaboration}
\label{app:collab}
\input{2021-07-05-Alice_Authorlist_2021-07-05.tex}  %%%%%%% done by webmaster team
\end{document}

%% file: Ch1_Introduction.tex
\section{Introduction}

In high-energy particle collisions, jet observables are sensitive to a variety of processes in quantum chromodynamics (QCD), from the initial hard (high $Q^2$) parton scattering to a scale evolution culminating in hadronization near $\Lambda_\text{QCD}$. Jets reconstructed with a radius (resolution) parameter near $R=1$ and with sufficiently large transverse momentum $\pTjet{}$ provide a proxy for the dynamics of the initial hard parton scattering, whereas those reconstructed with smaller $R$ or at lower $\pTjet{}$ become sensitive to nonperturbative effects. In this article, jet substructure observables are defined by clustering particles into a jet and then constructing an observable from its constituents to characterize its internal radiation pattern. 

Jet substructure techniques have provided one of the key tools to study rare event topologies in \pp{} collisions, for example by tagging boosted objects that decay into jets~\cite{Kogler_2019}.
Moreover, measurements of jet substructure enable stringent tests of perturbative QCD (pQCD) and facilitate studies of nonperturbative effects which are not yet under satisfactory theoretical control~\cite{Larkoski_2020}.
Jet substructure observables offer both flexibility and rigor: 
they can be constructed to be theoretically calculable from first-principles pQCD while simultaneously maintaining sensitivity to jet radiation in specific regions of phase-space.
Jet grooming algorithms, such as Soft Drop
\cite{Larkoski:2014wba, Dasgupta:2013ihk, Larkoski:2015lea}, can additionally be used to remove soft, wide-angle
radiation via well-controlled approaches, reducing nonperturbative effects.
This defines two families of jet substructure observables: one that can be constructed from all jet constituents and one based on a subset of jet constituents which remain after grooming procedures.

One such set of observables are the generalized jet angularities~\cite{Larkoski_2014, Almeida:2008yp}. Expanding upon the jet girth $g$ (also known as the jet radial moment), the generalized jet angularities form a class of jet substructure observables defined by

\begin{equation} \label{eq:1}
\lambda_{\alpha}^{\kappa} \equiv \sum_i z_i^{\kappa} \theta_i^{\alpha},
\end{equation}

where the sum runs over the jet constituents $i$,
and $\kappa$ and $\alpha$ are continuous free parameters.\footnote{
The notation $\lambda_{\alpha}$ is employed to represent the jet angularities instead of the commonly-used notation $\lambda_{\beta}$ in order to avoid conflict with the letter $\beta$, which is also used to denote the angular parameter of the Soft Drop grooming algorithm.}
The first factor $z_i \equiv \pTi/\pTjet$ describes the momentum fraction carried by
the constituent, and the second factor $\theta_i \equiv \Delta R_i / R$
denotes the separation in rapidity ($y$) and azimuthal angle
($\varphi$) of the constituent from the jet axis, where
$\Delta R_i \equiv \sqrt{\Delta y_i^2 + \Delta \varphi_i^2}$
and $R$ is the jet resolution parameter.
The jet angularities are infrared- and collinear- (IRC-)safe for 
$\kappa=1$ and $\alpha>0$~\cite{Berger_2003, Kang_2018}.
We consider the ungroomed jet angularities, denoted as
$\ang$, as well as the groomed jet angularities in which the sum runs
only over the constituents of the groomed jet, denoted as $\angsd$.
These include the jet girth~\cite{Catani:1992jc}, $\lambda_1$,
and the jet thrust~\cite{Farhi:1977sg}, $\lambda_2$,
which is related to the jet mass $m_{\mathrm{jet}}$ by 
$\lambda_2 = (m_{\mathrm{jet}}/\pTjet)^2 + \mathcal{O}(\lambda_2^2)$; $\lambda_2$, however, is more robust against nonperturbative effects
than $m_{\mathrm{jet}}$ since it does not depend explicitly on the hadron masses.

%This article presents the nominally infrared- and collinear- (IRC-)safe
%jet angularities, where $\kappa=1$ and $\alpha>0$~\cite{Berger_2003, Kang_2018}.
The IRC-safe jet angularities offer the possibility to systematically vary
the observable definition in a way that is theoretically calculable and
therefore provide a rich opportunity to study both perturbative and
nonperturbative QCD~\cite{Larkoski:2014tva, Procura:2018zpn, Stewart:2014nna, Lee_2007}.
This article considers jet angularities constructed from charged-particle jets. While charged-particle jets are IRC-unsafe~\cite{Chang:2013rca}, comparisons to these theoretical predictions can nonetheless be carried out by following a nonperturbative correction procedure, as outlined in Sec.~\ref{sect:calc}.
Jet angularities were recently calculated in \pp{} collisions 
both in the ungroomed~\cite{Kang_2018} and groomed~\cite{KANG201941}
cases, as well as for jets produced in association with a Z boson
\cite{Caletti:2021oor}.
These calculations use all-order resummation of large logarithms up to
next-to-leading-logarithmic (NLL$^\prime$) accuracy~\cite{Almeida:2014uva}.
Measurements of \ang{} and \angsd{} will serve to test these analytical
predictions, in particular the role of resummation effects and power
corrections. Moreover, by measuring multiple values of $\alpha$, one
can test the predicted scaling of nonperturbative shape functions that
are used to model hadronization, which depend only on a single
nonperturbative parameter for all values of $\alpha$
\cite{Korchemsky_1999_Fnp, Aschenauer_2020_Fnp}.

Several measurements of jet angularities have been performed in hadronic collisions.
The ungroomed jet angularity $\lambda_1$ has been measured in \pp{}
collisions by the ATLAS, CMS, and ALICE Collaborations~\cite{ang2018, Aad_2012, PhysRevD.98.092014} in addition to $\ppbar$ collisions by the CDF Collaboraiton~\cite{Aaltonen:2011pg}. The ungroomed jet angularity $\lambda_2$ has also been measured in \pp{} collisions by the CMS Collaboration~\cite{PhysRevD.98.092014}.
The closely related ungroomed and groomed jet mass have been extensively measured in \pp{} collisions by the ATLAS and CMS Collaborations~\cite{ATLAS:2012am, Aad_2012, Aad:2019wdr, PhysRevD.101.052007, PhysRevD.98.092014, Sirunyan:2019rfa, Sirunyan2018, Sirunyan_2017, CMS_dijet_mass_2013, CMS_pp_mass_2018, ATLAS-CONF-2018-014, ATLAS_SD_mass_2018}, 
and the ungroomed mass was also studied in $\ppbar$ collisions by the CDF Collaboration~\cite{Aaltonen:2011pg} and in \pPb{} collisions by the ALICE Collaboration~\cite{Acharya:2017goa}.
Many of these measurements have focused on using jet substructure
for tagging objects at high \pT{}, rather than for fundamental
studies of QCD, and with the exception of the jet mass 
there have not yet been comparisons of
jet angularities to analytical calculations,
nor have any such comparisons been made for charged-particle jets.
In this article, we
perform the first measurements of groomed jet angularities 
in \pp{} collisions, and a systematic scan of the IRC-safe ungroomed jet angularities.
These measurements focus on low to moderate \pTjet{}, and small to moderate $R$.
Moreover, the measurements are performed in \pp{} collisions at a center-of-mass energy
$\sqrt{s}=5.02$ TeV, the same center-of-mass energy at which ALICE recorded data in heavy-ion collisions during LHC Run 2,
and where no jet angularity measurements have been made.

These measurements serve as a baseline for future measurements of the jet angularities
in heavy-ion collisions, in which a deconfined state of
strongly-interacting matter is produced 
\cite{Jacak:2012dx, LHC1review, Braun-Munzinger:2015hba, TheBigPicture}. 
Measurements of jets and jet substructure in heavy-ion collisions 
may provide key insight into the physical properties of this deconfined state
\cite{ReviewXinNian, ReviewYacine, ReviewMajumder}.
The jet angularities are sensitive both to medium-induced broadening as well as
jet collimation~\cite{Yan:2020zrz, Elayavalli2017, Casalderrey-Solana:2019ubu};
by systematically varying the weight of collinear
radiation, one may be able to efficiently discriminate between jet quenching models.
In \PbPb{} collisions, $\lambda_1$ has been measured 
for $R=0.2$ by the ALICE Collaboration~\cite{ang2018},
and the ungroomed and groomed jet mass
have been measured for $R=0.4$ by the ATLAS, CMS, and ALICE Collaborations~\cite{Acharya:2017goa, Sirunyan2018, ATLAS-CONF-2018-014}.
The interpretation of previous measurements is unclear, with strong
modification being observed in \PbPb{} collisions compared to
\pp{} collisions for the case when $\alpha=1$ and $R=0.2$,
but little to no modification seen for the $R=0.4$ jet mass.
Future measurements over a range of $R$ and $\alpha$ offer
a compelling opportunity to disentangle the roles of medium-induced
broadening, jet collimation, and medium response in jet evolution.
By measuring small to moderate $R$ jets in \pp{} collisions, 
which are theoretically challenging and involve significant resummation effects~\cite{Acharya:2019jyg},
the ability of pQCD to describe the small-radius jets 
that are measured in heavy-ion collisions can be tested.

This article reports measurements of ungroomed and groomed jet
angularities for $\alpha=1$, 1.5, 2, and 3 in \pp{} collisions at
{$\sqrt{s}=5.02$ TeV}. In addition to the standard jet girth ($\alpha=1$)
and jet mass (related to $\alpha=2$) parameters, $\alpha=1.5$ and $\alpha=3$ are included to test the universality
of a nonperturbative shape function by varying effects of soft,
wide-angle radiation, as discussed below in Sec.~\ref{sect:shape_fn},
and to serve as a reference for future jet quenching measurements
in heavy-ion collisions.
Grooming is performed according to the Soft Drop grooming procedure
with $\zcut=0.2$ and $\betasd=0$~\cite{Mulligan:2020tim}.
Charged particle jets were reconstructed at midrapidity 
using the anti-$k_\mathrm{T}$ algorithm with jet resolution (radius)
parameters ${R=0.2}$ and ${R=0.4}$
in four equally-sized \pTchjet{} intervals from 20 to 100 \GeVc{}. 
The results are compared to NLL$^\prime$ pQCD predictions,
as well as to the PYTHIA8~\cite{pythia} and Herwig7~\cite{B_hr_2008,Bellm:2015jjp} Monte Carlo generators.

%% file: Ch2_ExperimentalSetup.tex
\section{Experimental setup and data sets}

A description of the ALICE detector and its performance can be found in
Refs.~\cite{aliceDetector, alicePerformance}. The \pp{} data used in this
analysis were collected in 2017 during LHC Run 2 at $\sqrt{s} = 5.02$ TeV~\cite{LHCmachine}.
A minimum bias (MB) trigger was used; this requires a coincidence of hits in the V0 scintillator detectors, which provide full azimuthal coverage and cover the pseudorapidity ranges of $2.8 < \eta < 5.1$ and $-3.7 < \eta < -1.7$~\cite{Abbas:2013taa}.
The event selection also requires the location of the primary vertex to be within $\pm10$ cm from the nominal interaction point (IP) along the beam direction and within 1 cm of the IP in the transverse plane. 
Beam-induced background events were removed using two neutron
Zero Degree Calorimeters located at $\pm112.5$ m along the beam axis 
from the center of the detector.
Events with multiple reconstructed vertices were rejected, and track quality selection criteria ensured that tracks used in the analysis were from only one vertex.
Events were acquired at instantaneous luminosities between approximately
$10^{30}$ and $10^{31} \;\rm{cm}^{-2}\rm{s}^{-1}$,
corresponding to a low level of pileup with approximately
$0.004 < \mu < 0.03$ events per bunch crossing.
The pp data sample contains 870 million events and corresponds to
an integrated luminosity of 18.0(4) nb$^{-1}$~\cite{ppXsec}.

This analysis uses charged particle tracks reconstructed from clusters in both the 
Time Projection Chamber (TPC)~\cite{Alme_2010} 
and the Inner Tracking System (ITS)~\cite{Aamodt:2010aa}. 
Two types of tracks are defined: global tracks and complementary tracks.
Global tracks are required to include at least
one hit in the silicon pixel detector (SPD), comprising the first
two layers of the ITS, and to satisfy a number of quality criteria~\cite{Acharya:2019tku}, including 
having at least 70 out of a maximum of 159 TPC space points and at least 80\% of the geometrically findable space points in the TPC.
Complementary tracks do not contain any hits in the SPD, but
otherwise satisfy the tracking criteria, and are refit with a constraint to the primary vertex of the event.
Including this second class of tracks ensures approximately uniform
azimuthal acceptance, while preserving similar transverse momentum \pT{} resolution to tracks with SPD hits, as determined from the fit quality.
Tracks with $\pTtrack > 0.15 \;\GeVc$ are 
accepted over pseudorapidity $|\eta| < 0.9$ and azimuthal angle 
$0 < \varphi < 2\pi$.
All tracks are assigned a mass equal to the $\pi^{\pm}$ mass.

The instrumental performance of the ALICE detector and its response to particles is estimated with a GEANT3~\cite{Brun:1119728} model. 
The tracking efficiency in \pp{} collisions, as estimated by propagating pp events from PYTHIA8 Monash 2013~\cite{pythia} through the ALICE GEANT3 detector simulation, is approximately 67\% 
at $\pTtrack=0.15 \;\GeVc$, rises to approximately 84\% at $\pTtrack=1 \;\GeVc$, and remains above 75\% at higher \pT.
The momentum resolution $\sigma(\pT)/\pT$ is estimated from the covariance matrix of the track fit~\cite{alicePerformance}
and is approximately 1\% at $\pTtrack=1$ \GeVc{}. This increases with \pTtrack{}, reaching approximately 4\% at $\pTtrack=50$ \GeVc{}.

%% file: Ch3_DataAnalysis.tex
\section{Analysis method}

%%%%%%%%%%%%%%%%%%%%%%%%%%%%%%%%%%%%%%%%%%%%%%%%%%%%
\subsection{Jet reconstruction}

Jets are reconstructed from charged tracks with $\pT{} > 150$ MeV$/c$ using the FastJet package~\cite{Cacciari:2011ma}. 
The anti-$k_{\mathrm{T}}$ algorithm is used with the $E$ recombination scheme
for resolution parameters $R=0.2$ and $0.4$~\cite{antikt}. 
All reconstructed charged-particle jets in the transverse momentum range 
$5 < \pTchjet{} < 200$ GeV$/c$ are analyzed in order to maximize statistics in the unfolding procedure (described below). Each jet axis is required to be within the fiducial volume of the TPC,
$\left| \etajet \right| < 0.9 - R$.
Jets containing a track with $\pT>100\;\GeVc$ are removed from
the collected data sample, due to limited momentum resolution.
In order to make consistent comparisons between the data and the theoretical
calculations, the background due to the underlying event is not subtracted
from the data, and instead the underlying event (along with other nonperturbative
effects) is included in model corrections, as described in Sec.~\ref{sect:calc}.

The jet reconstruction performance is studied by
comparing jets reconstructed from PYTHIA8-generated events at ``truth level'' (before the particles undergo interactions 
with the detector) to those at ``detector level'' (after the ALICE GEANT3 detector simulation). Two collections of jets are constructed: pp truth level (PYTHIA truth) and
pp detector level (PYTHIA with detector simulation). The detector-level jets are then geometrically matched with truth-level jets within $\Delta R < 0.6$ $R$ while additionally requiring that each match be unique.
Table~\ref{table:jetreco} shows approximate values of the mean jet energy scale shift,
$\Delta_{\mathrm{JES}}=\left< \left( \pTdet - \pTtruth \right) / \pTtruth \right>$, 
the jet energy resolution, 
$\mathrm{JER}=\sigma\left(\pTdet \right) \left/ \pTtruth \right.$, 
and the jet reconstruction efficiency, 
$\varepsilon_{\mathrm{reco}}$, for both $R=0.2$ and $R=0.4$,
where \pTdet{} is the detector-level \pTchjet{}, and \pTtruth{} is the truth-level \pTchjet{}.
The jet energy scale shift is a long-tailed asymmetric distribution due to tracking inefficiency~\cite{Abelev:2013kqa} with a peak at $\pTdet = \pTtruth$, 
and $\Delta_{\mathrm{JES}}$ should be understood only as a 
rough characterization of this distribution.

\begin{table}[!t]
\centering
\caption{Approximate values characterizing the jet reconstruction performance for $R=0.2$ and $0.4$ in \pp{} collisions. $\Delta_\text{JES}$ is the mean jet energy scale shift, $\text{JER}$ is the jet energy resolution, and $\epsilon_\text{reco}$ is the reconstruction efficiency.}
\begin{tabular}{ l ccccc }
\tabularnewline \hline \hline & \multicolumn{2}{c}{$R=0.2$} & \multicolumn{2}{c}{$R=0.4$}
\tabularnewline \hline \pTchjet{} & $20 \;\GeVc$ & $100\;\GeVc$ & $20\;\GeVc$ & $100\;\GeVc$
\tabularnewline \hline \quad $\Delta_{\mathrm{JES}}$ & --12\% & --24\% & --13\% & --21\%
\tabularnewline \quad $\mathrm{JER}$ & 22\% & 21\% & 21\% & 21\%
\tabularnewline \quad $\varepsilon_{\mathrm{reco}}$ & 94\% & 100\% & 97\% & 100\

\tabularnewline \hline 
\end{tabular}
\label{table:jetreco}
\end{table}

The ungroomed jet angularities are reconstructed using all of the 
charged-particle jet constituents according to Eq. (\ref{eq:1}).
For the groomed jet angularities,
Soft Drop grooming~\cite{Larkoski:2014wba} is performed, in which the constituents of each jet are reclustered with the Cambridge--Aachen algorithm~\cite{Dokshitzer_1997} 
with resolution parameter $R$, forming an angularly-ordered tree data structure.
Each node corresponds to a constituent track, and each edge is a branch splitting defined by
$z \equiv \frac{\pTsub}{\pTlead + \pTsub}$ and 
$\theta \equiv \frac{\Delta R}{R} \equiv \frac {\sqrt{\Delta y ^2 + \Delta \varphi ^2}}{R}$.
The jet tree is then traversed starting from the largest-angle splitting,
and the Soft Drop condition, $z > \zcut \theta^{\betasd}$, is
recursively evaluated. Here, $z$ is the subleading branch \pT{}
fraction defined above, and $\zcut$ and ${\betasd}$ are tunable, free
parameters of the grooming algorithm. For this analysis, 
$\betasd=0$ is used to maximize the perturbative calculability
\cite{KANG201941}, while $\zcut{} = 0.2$ is chosen (as
opposed to the more common $\zcut{} = 0.1$) since higher-accuracy
branch tagging can be achieved in future heavy-ion collision analyses
\cite{Mulligan:2020tim}.
If the Soft Drop condition is not satisfied, then the softer 
subleading branch is discarded and the next splitting in the harder
branch is examined in the same way. If, however, the condition is
satisfied, then the grooming procedure is concluded, with all
remaining constituents defining the groomed jet. 
The groomed jet angularity is then defined according to
Eq. (\ref{eq:1}) using the groomed jet constituents, but still with
the ungroomed \pTchjet{} and ungroomed jet axis to define $\theta_i$,
since the groomed jet observable is a property of the original
(ungroomed) jet object. Note that while the ungroomed \pTchjet{} is
IRC-safe, the groomed $p_\text{T,g}^\text{ch jet}$ is Sudakov safe~\cite{Cal_2020}.
If the jet does not contain a splitting that passes the Soft Drop
condition, then the groomed jet contains zero constituents
(``untagged'') and does not have a defined groomed jet angularity.

%%%%%%%%%%%%%%%%%%%%%%%%%%%%%%%%%%%%%%%%%%%%%%%%%%%%%%%%%%%%%
%%%%%%%%%%%%%%%%%%%%%%%%%%%%%%%%%%%%%%%%%%%%%%%%%%%%%%%%%%%%%
\subsection{Corrections}

The reconstructed \pTchjet{} and \ang{} differ from
their true values due to tracking inefficiency,
particle--material interactions, and track \pT{} resolution.
To account for these effects, PYTHIA8 Monash 2013 
\cite{pythia, Skands_2014} and the ALICE GEANT3 detector
simulation are used to construct a 4D response matrix
that describes the detector response mapping of $\pTtruth$ and
$\angtruth$ to $\pTdet$ and $\angdet$, where \pTdet{} and
\pTtruth{} are as above, and $\angdet$ and $\angtruth$ are
the analogous detector- and truth-level \ang{}. 
The truth-level jet was constructed from the charged primary
particles of the PYTHIA event, defined as all particles with a 
mean proper lifetime larger than 1 cm/$c$, and excluding the decay
products of these particles~\cite{primaryParticleALICE}.

A 2D unfolding in \pTchjet{} and \ang{} is then performed using
the iterative Bayesian unfolding algorithm
\cite{DAgostini,dagostini2010improved} implemented in 
the RooUnfold package~\cite{roounfold} to recover the
true jet spectrum at the charged-hadron level. This technique
utilizes a ``prior" distribution (equivalent to the per-bin MC
prediction) as a starting point, before iteratively updating the
distribution using Bayes' theorem in conjunction with the calculated
response matrix and measured data (see Refs.
\cite{DAgostini,dagostini2010improved} for details). Since the jet
yield in each reported \pTchjet{} interval varies widely, with
higher-\pTchjet{} jets being less probable than lower-\pTchjet{} jets,
and since the shape and mean value of the jet angularity distributions
also changes with \pTchjet{}, a separate 2D unfolding for each reported
\pTchjet{} bin is performed in order to optimize the observable binning
at both truth and detector levels, thus ensuring sufficient jet yield
is included in the procedure for all distributions while
simultaneously maximizing the number of bins for regions of
phase space where higher yield is available. The bin migration
in all cases is dominated by a strong diagonal mapping in the
response matrix coupled with a slight smearing along the \pTtruth{}
and \angtruth{} axes. The smearing in \ang{} is roughly symmetric
about the diagonal, whereas the smearing in \pTchjet{} tends to be skewed
towards lower values of \pTtruth{} due to tracking efficiency effects.

In the groomed case, the number of untagged jets in 
the unfolding procedure is included as an additional bin adjacent
to the lower edge of the \ang{} distributions. This is done so
that the unfolding procedure will correct for detector effects on
the groomed jet tagging fraction as well as account for
bin migration effects for jets which are groomed
away at detector-level but not truth-level, or vice versa.

To validate the performance of the unfolding procedure, a set of
refolding and closure tests is performed, in which either the response
matrix is multiplied by the unfolded data and compared to the original
detector-level spectrum, or in which the shape of the input MC spectrum
is modified to account for the fact that the actual distribution may
be different than the MC input spectrum. 
The number of iterations, which sets the strength of regularization,
is chosen to be the minimal value such that all unfolding tests succeed.
This results in the number of iterations being equal to 3 for all distributions. 
In all cases, closure is achieved compatible with statistical uncertainties.

The distributions after unfolding are corrected for the kinematic
efficiency, defined as the efficiency of reconstructing a truth-level jet at a particular \pTtruth{} and \angtruth{} value given a reconstructed jet \pTdet{} and \angdet{} range. Kinematic
inefficiency results from effects including smearing from
the Soft Drop threshold and \pT{}-smearing of the jet out of the selected \pTdet{} range. Any ``missed'' jets, those jets which exist at truth level but not
at detector level, are handled by this kinematic efficiency correction.
In this analysis, minimal detector-level cuts are applied, 
and the kinematic efficiency is therefore greater than $99\%$ in all cases.
Since a wide \pTtruth{} range is taken, the effect of ``fake'' jets,
those jets which exist at detector level but not truth level,
is taken to be negligible.

%% file: Ch4_SystematicUncertainties.tex
\section{Systematic uncertainties}
%%%%%%%%%%%%%%%%%%%%%%%%%%%%%%%%%%%%%%%%%%%%%%%%%%%%%%%%%%%%%
%%%%%%%%%%%%%%%%%%%%%%%%%%%%%%%%%%%%%%%%%%%%%%%%%%%%%%%%%%%%%

The systematic uncertainties 
in the unfolded results arise from uncertainties in
the tracking efficiency and unfolding procedure, as well as
the model-dependence of the response matrix,
and the track mass assumption.
Table~\ref{tab:sysunc} summarizes the systematic uncertainty contributions.
Each of these sources of uncertainty dominate in certain 
regions of the measured observables, 
with the exception of the track mass assumption which is small
in all cases.
The total systematic uncertainty is taken as the sum in quadrature of 
the individual uncertainties described below.

%%%%%%%%%%%%%%%%%%%%%%%%%%%%%%%%%%%%%%%%%%%%%%%%%%%%%%%%%%%%%
%\subsubsection{Tracking efficiency}

The tracking efficiency uncertainty is estimated to be 4\% 
by varying track selection parameters and the ITS--TPC
matching requirement. In order to assign a systematic uncertainty
to the nominal result, a response matrix is constructed using the same
techniques as for the final result except that an additional 4\%
of tracks are randomly rejected before the jet finding. 
This response matrix is then used to unfold the distribution in
place of the nominal response matrix, and the result is compared to
the default result, with the differences in each bin taken as a
symmetric uncertainty. This uncertainty
constitutes a smaller effect in the groomed jet angularities,
where single-particle jets, being the most sensitive to the
tracking efficiency, are groomed away by the Soft Drop condition.
The uncertainty on the track momentum resolution is a sub-leading
effect to the tracking efficiency and is taken to be negligible.

%%%%%%%%%%%%%%%%%%%%%%%%%%%%%%%%%%%%%%%%%%%%%%%%%%%%%%%%%%%%%
%\subsubsection{Unfolding uncertainties}

Several variations of the unfolding procedure are performed
in order to estimate the systematic
uncertainty arising from the unfolding regularization procedure:

\begin{enumerate}
    \item The number of iterations 
    was varied by $\pm2$
    and the average difference
    with respect to the nominal result is taken as the systematic uncertainty.
    \item The prior distribution is scaled by a power law in \pTchjet{} 
    and a linear scaling in \ang{}, $(\pTchjet)^{\pm0.5} \times [1\pm(\ang-0.5)]$. 
    The average difference between the result unfolded with this prior and the original 
    is taken as the systematic uncertainty. 
    \item The binning in \ang{} was varied to be slightly
    finer and coarser than the nominal binning, by combining (splitting) some adjacent bins with low (high) jet yield, or by shifting the bin boundaries to be between the nominal boundaries.
    \item The lower and upper bounds in the \pTdet{} range were increased to 10 and decreased to 120 \GeVc{}, respectively. These values are chosen as reasonable values to estimate sensitivity to truncation effects.
\end{enumerate}

The total unfolding systematic uncertainty is then the standard deviation of the variations, $\sqrt{\sum_{i=1}^{N} \sigma_{i}^{2} / N}$, 
where $N=4$ and $\sigma_{i}$ is the systematic uncertainty due to a single variation, since they each comprise independent measurements of the same underlying systematic uncertainty in the regularization.

% %%%%%%%%%%%%%%%%%%%%%%%%%%%%%%%%%%%%%%%%%%%%%%%%%%%%%%%%%%%%%
% \subsubsection{Model dependence}
A systematic uncertainty associated with the model-dependent 
reliance on the Monte Carlo generator which is used to unfold the
spectra is included. We construct a fast simulation to parameterize
the tracking efficiency and track \pT{} resolution, and build
response matrices using PYTHIA8 Monash 2013 and Herwig7 (default tune)
as generators. Even though a full detector simulation using PYTHIA8
has also been generated, a fast simulation is used for this purpose
so that there is complete parity between the two generators in the
calculation of this systematic uncertainty. This fast simulation
provides agreement within $\pm 10\%$ of the full detector simulation
for $R=0.2$ jets, with some larger deviations seen in the tails of the
jet angularity distributions for $R=0.4$ jets.
These two response matrices are then used to unfold the measured
data, and the differences between the two unfolded results in each
interval are taken as a symmetric uncertainty. This uncertainty is
most significant at lower \pTchjet{}.

% %%%%%%%%%%%%%%%%%%%%%%%%%%%%%%%%%%%%%%%%%%%%%%%%%%%%%%%%%%%%%
% \subsubsection{Track mass assumption}
In order to assess the uncertainty due to the track mass assumption,
K$^\pm$ meson and proton masses are randomly assigned to 13\% and 5.5\%
of tracks, respectively, in both the data and the response matrix.
These numbers are chosen from the (approximate) inclusive number of
each respective particle measured at midrapidity in \pp{} events by
ALICE~\cite{protonKpi_ALICE}. Neither the measurement inside the jets nor
the \pTchjet{}-dependence are considered, so these numbers are taken
to constitute a reasonable maximum uncertainty. The bin-by-bin
difference of the unfolded result to the nominal result is taken
as a symmetric uncertainty.

%%%%%%%%%%%%%%%%%%%%%%%%%%%%%%%%%%%%%%%%%%%%%%%%%%%%%%%%%%%%%%%%%%%%%%%%%%

\begin{table}[]
\centering
\caption{Summary of systematic uncertainties for a representative sample of $\alpha$, $R$, and $\pTchjet$. A moderately high $60 < \pTchjet{} < 80$ \GeVc{} with $R=0.4$ is chosen to show the variation with $\alpha$, and two additional rows show the trends with smaller \pTchjet{} and $R$.}
\begin{tabular}{ l cccccccc }
\tabularnewline \hline \hline & & & & \multicolumn{5}{c}{Relative uncertainty}
\tabularnewline \cline{5-9} & $\alpha$ & $R$ & \pTchjet (\GeVc) & Trk. eff. & Unfolding  & Generator & Mass hypothesis & Total
%\vspace{0.5mm}
\tabularnewline \hline \ang \\
& 1 & 0.4 & 60--80  & 1--15\%  & 2--7\% &  1--5\%   & 0--2\%   & 7--16\%  \\
& 2 & 0.4 & 60--80  & 1--10\%  & 1--8\% &  1--5\%   & 1--3\%   & 4--12\%  \\
& 3 & 0.4 & 60--80  & 1--10\%  & 2--4\% &  1--4\%   & 0--4\%   & 4--11\%  \\
& 2 & 0.4 & 20--40  & 1--16\%  & 1--4\% &  1--43\%  & 0--5\%   & 2--44\%  \\
& 2 & 0.2 & 60--80  & 2--12\%  & 2--7\% &  1--9\%   & 0--2\%   & 3--12\%  \\
\hline \angsd \\
& 1 & 0.4 & 60--80  & 1--7\%   & 2--8\% &  1--6\%   & 0--4\%   & 2--13\%  \\
& 2 & 0.4 & 60--80  & 1--8\%   & 2--9\% &  1--5\%   & 0--4\%   & 3--12\%  \\
& 3 & 0.4 & 60--80  & 1--6\%   & 2--7\% &  1--11\%  & 0--7\%   & 4--16\%  \\
& 2 & 0.4 & 20--40  & 1--8\%   & 2--5\% &  1--40\%  & 0--3\%   & 2--42\%  \\
& 2 & 0.2 & 60--80  & 1--7\%   & 1--8\% &  1--12\%  & 0--3\%   & 1--15\%  \\
\hline \hline
\end{tabular} 
\label{tab:sysunc}
\end{table}

%% file: Ch5_Results.tex
\section{Results and discussion}

%%%%%%%%%%%%%%%%%%%%%%%%%%%%%%%%%%%%%%%%%%%%%%%%%%%%%%%%%%%%%
%%%%%%%%%%%%%%%%%%%%%%%%%%%%%%%%%%%%%%%%%%%%%%%%%%%%%%%%%%%%%

We report the $\ang$ and $\angsd$ distributions
for $\alpha=1$, 1.5, 2, and $3$ in four equally-sized intervals of \pTchjet{} between 20 and 100 GeV/c.
The distributions are reported as differential cross sections:
\begin{equation} \label{eq:7}
\frac{1}{\sigma} \frac{\rm{d}\sigma}{\rm{d}\ang}
\equiv
\frac{1}{N_{\text{jets}}} \frac{\mathrm{d}N_\text{jets}} {\rm{d}\ang}
\text{ (ungroomed),} \hspace{1em} \rm{or} \hspace{1em}
\frac{1}{\sigma_{\rm{inc}}} \frac{\rm{d}\sigma}{\rm{d}\angsd}
\equiv
\frac{1}{\it{N}_{\text{inc jets}}}
\frac{\mathrm{d}\it{N}_\text{gr jets}} {\rm{d}\angsd} \text{ (groomed),}
\end{equation}
where $N_{\rm{jets}}$ is the number of jets within a given \pTchjet{}
range and $\sigma$ is the corresponding cross section. For the groomed case, some jets are removed by the grooming procedure, and therefore two different quantities are defined: $\it{N}_\text{gr jets}$, the number of jets which have at least one splitting satisfying the Soft Drop condition, and $\it{N}_{\text{inc jets}}$, the total number of inclusive jets, with both $\it{N}_\text{gr jets}$ and $\it{N}_{\text{inc jets}}$ being within the given \pTchjet{} range. $\sigma_{\rm{inc}}$ is the cross section corresponding to the latter inclusive quantity. For the ungroomed case, $\it{N}_{\text{inc jets}} = N_{\rm{jets}}$ and $\sigma = \sigma_{\rm{inc}}$, so the redundant labels are dropped. It is useful to normalize the groomed differential cross section by the number of inclusive jets since the groomed jet angularities are a property of the inclusively-measured jet population and are thus typically normalized as such in theoretical calculations~\cite{KANG201941}.

The ungroomed jet angularity distributions are shown in Fig.
\ref{fig:ang-1} and Fig.~\ref{fig:ang-2} for $R=0.4$ and $R=0.2$,
respectively. By the definitions given in Eq.~\ref{eq:7},
these distributions are all normalized to unity. As $\alpha$
increases, the distributions skew towards
small $\ang$, since $\theta_i$ is smaller than unity. 
For larger $R$, the distributions are narrower than for smaller $R$, 
as expected due to the collinear nature of jet fragmentation.
For small $R$ and low \pTchjet{} there is a visible peak at $\ang=0$,
which is due to single particle jets. These distributions are compared to
PYTHIA8 Monash 2013~\cite{pythia, Skands_2014} and
Herwig7 (default tune)~\cite{B_hr_2008,Bellm:2015jjp}
from truth-level projections of the respective response matrices, with jet
reconstruction assigning tracks the $\pi^\pm$ meson mass as in the measured data. 
These comparisons show deviations up to approximately $+50\% (-30\%)$. 
The largest deviations are for small values of \ang{}, where nonperturbative
physics becomes significant (see Sec.~\ref{sect:calc} for discussion).

The groomed jet angularity distributions for $\zcut=0.2$ and
$\betasd=0$ are shown in Fig.~\ref{fig:angsd-1}
for $R=0.4$ and Fig.~\ref{fig:angsd-2} for $R=0.2$.
Note that these distributions are shown on a logarithmic scale
due to the distributions being more strongly peaked and falling
faster with \ang{} as compared to the ungroomed distributions.
The groomed jet angularities have significantly smaller values 
than the ungroomed jet angularities, due to the
removal of soft wide-angle radiation.
The fraction of ``untagged'' jets, those that do not contain a
splitting which passes the Soft Drop condition, ranges from 10 to 12\%.
Unlike the ungroomed jet angularities, which are normalized to unity,
the groomed jet angularities are normalized to the Soft Drop tagging fraction.
Since the tagging rate is fairly large, the measured
distributions are therefore normalized close to unity.
PYTHIA and Herwig describe the groomed jet angularities slightly better
than the ungroomed jet angularities, with most deviations seen in the
ungroomed distributions improving by 10--20\% in the groomed case.
Comparing to the two MC generators, the data are in slightly
better agreement with Herwig7 than with PYTHIA8, especially for $R=0.4$.

\afterpage{%
\begin{figure}[!hb]
\centering{}
\vspace{5em}
\includegraphics[scale=0.8]{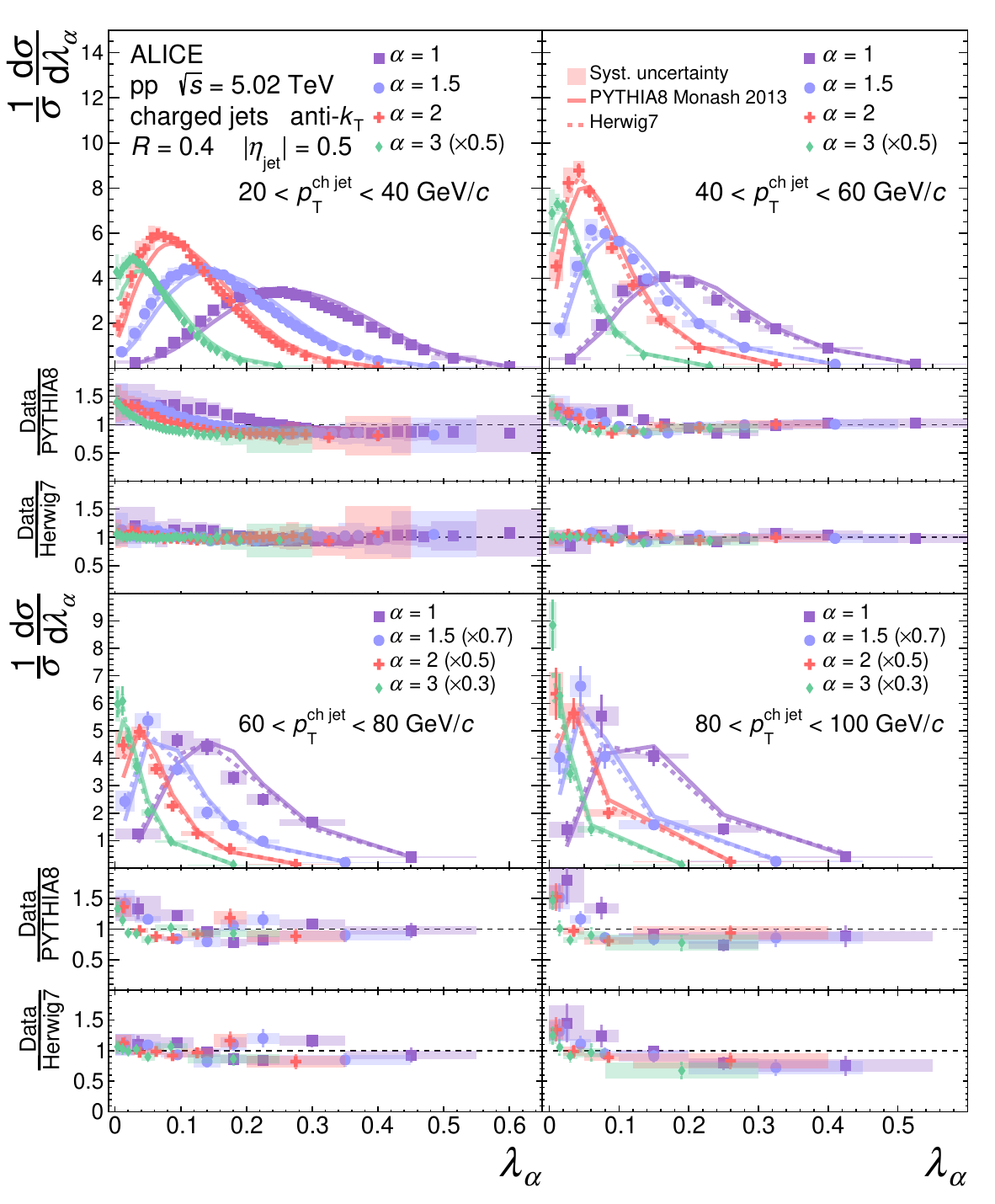}
\caption{Comparison of ungroomed jet angularities \ang{} in \pp{}
collisions for $R=0.4$ to MC predictions using PYTHIA8 and Herwig7, as described in the text. Four equally-sized \pTchjet{} intervals are shown, with edges ranging between 20 and 100 \GeVc{}. The distributions are normalized to unity.}
\label{fig:ang-1}
\end{figure}
\clearpage

\begin{figure}[!hb]
\centering{}
\vspace{5em}
\includegraphics[scale=0.8]{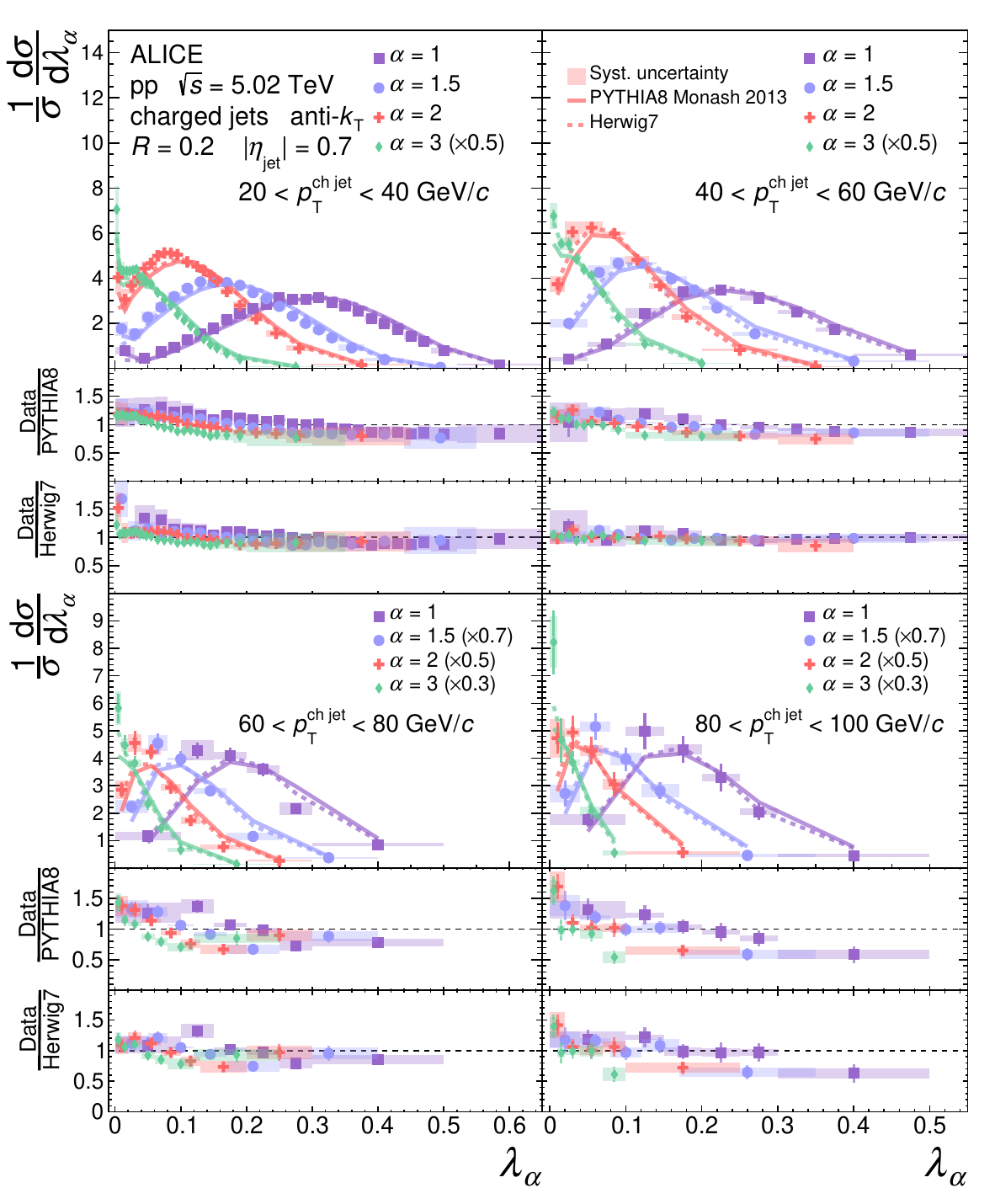}
\caption{Comparison of ungroomed jet angularities \ang{} in \pp{}
collisions for $R=0.2$ to MC predictions using PYTHIA8 and Herwig7, as described in the text. Four equally-sized \pTchjet{} intervals are shown, with edges ranging between 20 and 100 \GeVc{}. The distributions are normalized to unity.}
\label{fig:ang-2}
\end{figure}
\clearpage

\begin{figure}[!hb]
\centering{}
\vspace{5em}
\includegraphics[scale=0.8]{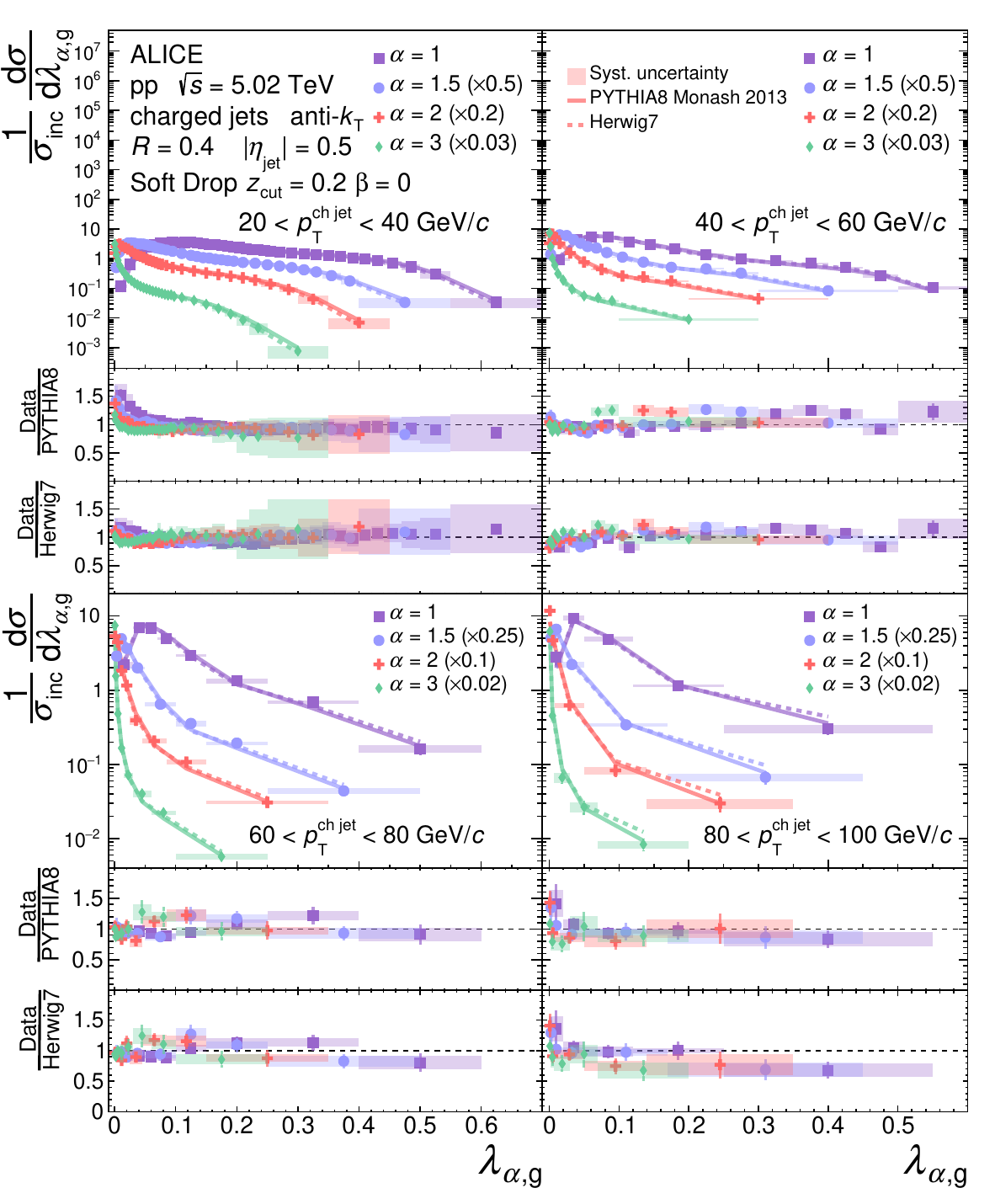}
\caption{Comparison of groomed jet angularities \angsd{} in \pp{}
collisions for $R=0.4$ to MC predictions using PYTHIA8 and Herwig7, as described in the text. Four equally-sized \pTchjet{} intervals are shown between 20 and 100 \GeVc{}. The distributions are
normalized to the groomed jet tagging fraction.}
\label{fig:angsd-1}
\end{figure}
\clearpage

\begin{figure}[!hb]
\centering{}
\vspace{5em}
\includegraphics[scale=0.8]{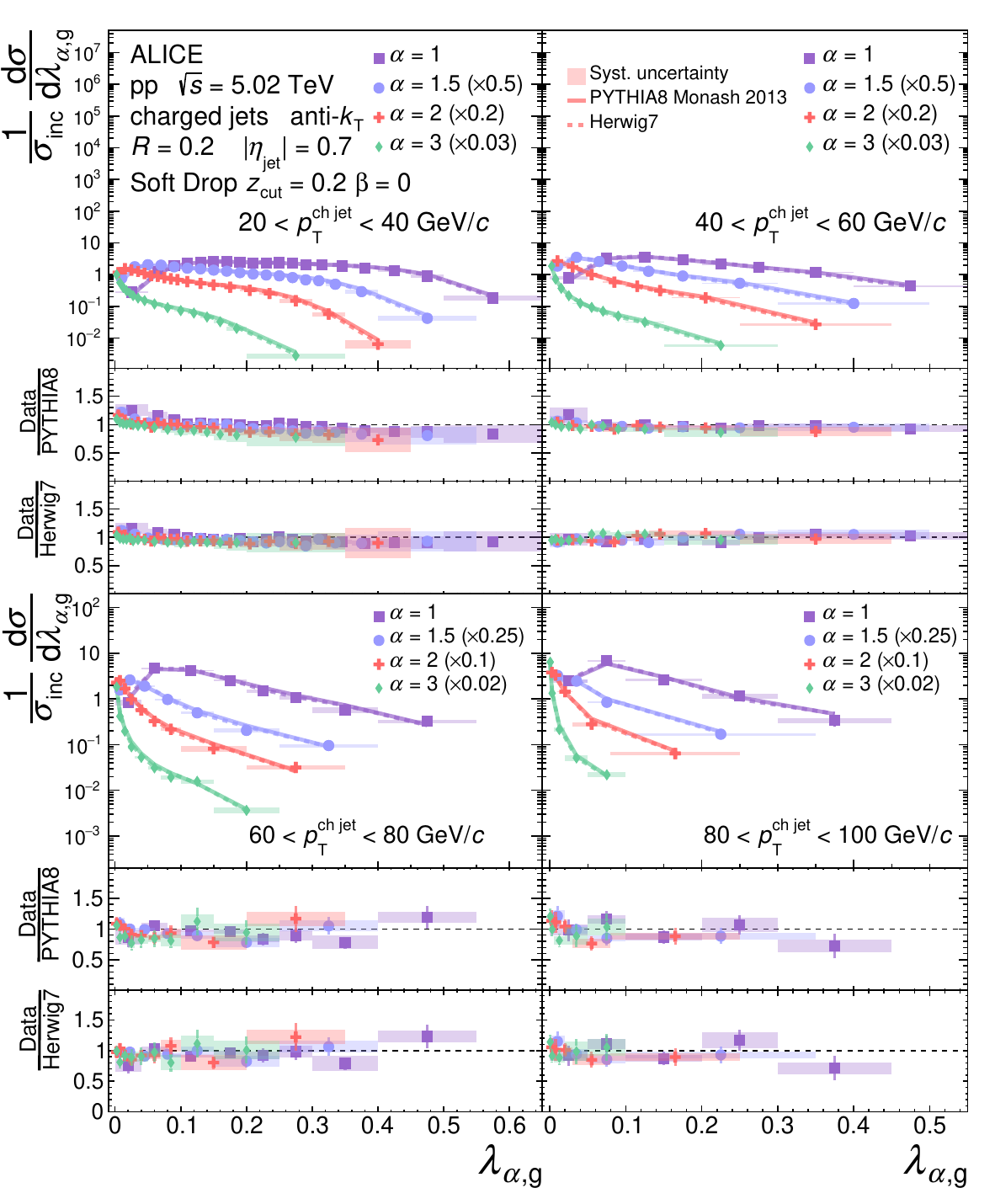}
\caption{Comparison of groomed jet angularities \angsd{} in \pp{}
collisions for $R=0.2$ to MC predictions using PYTHIA8 and Herwig7, as described in the text. Four equally-sized \pTchjet{} intervals are shown between 20 and 100 \GeVc{}. The distributions are
normalized to the groomed jet tagging fraction.}
\label{fig:angsd-2}
\end{figure}
\clearpage
}

The data cover a wide range of $\alpha$ and multiple $R$
down to low \pT{}, and therefore are subject to varying influence
from nonperturbative effects. Accordingly, these data can be used to 
study nonpertubative effects. The level and location of the disagreements
with PYTHIA and Herwig provide further constraints on 
nonperturbative effects in MC event generators.
Moreover, the comparison of the groomed and the ungroomed jet
angularities with MC event generators allows direct sensitivity to
radiation that was groomed away, which is highly nonperturbative.

%%%%%%%%%%%%%%%%%%%%%%%%%%%%%%%%%%%%%%%%%%%%%%%%%%%%%%%%%%%%%
%%%%%%%%%%%%%%%%%%%%%%%%%%%%%%%%%%%%%%%%%%%%%%%%%%%%%%%%%%%%%
\subsection{Comparison to analytical calculations}
\label{sect:calc}

The measured ungroomed and groomed jet angularities are compared 
with analytical calculations~\cite{Kang_2018, KANG201941} which use
all-order resummations of large logarithms to next-to-leading
logarithmic (NLL$^\prime$) accuracy~\cite{Almeida:2014uva}.
In particular, the calculations resum logarithms of \ang{}, $R$, and
\zcut. In the case of the \ang{} logarithms, the cumulant of the
cross section includes the complete set of terms of form
$\alpha_\mathrm{s}^n \ln^k \ang$ for $k=2n$, $2n-1$, and $2n-2$.
The calculations are valid up to power corrections in \ang{}, $R$,
and \zcut, and do not include non-global logarithms
\cite{Dasgupta:2001sh}.
These calculations are based on the framework of Soft Collinear
Effective Theory (SCET)~\cite{SCETplaceholder}, 
in which the jet cross section is factorized into
a ``hard function" corresponding to the initial scattering, and a
``jet function" corresponding to the fragmentation of a
hard-scattered parton into a jet.
For the calculation of the jet angularities, the jet function is
then further factorized into collinear and soft functions.
Systematic uncertainties on the analytical predictions 
are estimated by systematically varying fifteen combinations of
scales that emerge in the calculation.

For the ungroomed jet angularities, the collinear-soft momentum scale for the factorization formalism
becomes nonperturbative for~\cite{Kang_2018}
\begin{equation} \label{eq:3}
\ang \lesssim \frac{\Lambda}{\pTchjet R},
\end{equation}
where $\Lambda$ is the energy scale at which $\alpha_s$ becomes
nonperturbative, which is taken to be approximately $1\;\GeVc$.
For the groomed jet angularities with $\betasd=0$, this soft factorization scale becomes nonperturbative for~\cite{KANG201941}
\begin{equation} \label{eq:4}
\angsd \lesssim 
\zcut^{1-\alpha} \bigg( \frac{\Lambda}{\pTchjet{} \;R} \bigg)^{\alpha}.
\end{equation}
Accordingly, the analytical predictions are expected to describe the
data only at sufficiently large \ang, which depends on $\pTchjet$,
$R$, and $\zcut$. 
On the other hand, for $\ang = \mathcal{O}(1)$, 
power corrections in \ang{} become important,
and are not included in the NLL$^\prime$ calculations.
Note that for $\angsd{} > z_\text{cut}$, the groomed
and ungroomed predictions are identical at the parton level.

For values of \ang{} that are sufficiently large to be described by
SCET, corrections for nonperturbative effects must still be applied in
order to compare these parton-level calculations to our
charged-hadron-level measurements. These nonperturbative effects
include hadronization, the underlying event, and the selection
of charged particle jets. Note that track-based observables
are IRC-unsafe. In general, nonperturbative track functions can be
used to directly compare track-based measurements to analytical
calculations~\cite{Chang:2013rca, Chen:2020vvp, Chien:2020hzh}; 
however, such an approach has not yet been developed for jet
angularities. 
Two techniques are used, described in the following subsections,
to apply the nonperturbative corrections. 

%%%%%%%%%%%%%%%%%%%%%%%%%%%%%%%%%%%%%%%%%%%%%%%%
\subsubsection{MC-based hadronization correction}

The first technique relies solely on MC generators to transform
the parton-level calculations into the final predictions at the charged-hadron level.
Two response matrices are constructed, one using PYTHIA 8.244 and the
other using Herwig7, which map the jet angularity distributions
from jets reconstructed at the final-state
parton level (after the parton shower) to those from jets
reconstructed at the charged-hadron level. This is done by requiring
a unique geometrical match between the parton and charged-hadron-level
jets of $\Delta R < R/2$. The PYTHIA8 simulation uses the default
Monash 2013 tune, which is tuned to both e$^+$e$^-$ and p$\bar{\text{p}}$ data
\cite{Skands_2014}, with the only change being that the minimum
shower \pT{} (\texttt{TimeShower:pTmin}) is set to 0.2 \GeVc{}, one half of
its default value, in order to better match the NLL$^\prime$ predictions at parton level.
Herwig7 is also run with the default tune~\cite{Gieseke:2012ft}. The
response matrix generated with both MC simulations is 4D,
mapping $p_\text{T}^\text{parton jet}$
and $\lambda_\beta^\text{parton jet}$ to $\pTtruth$ and $\angtruth$.

Since the NLL$^\prime$ predictions are generated as normalized
distributions, each \pTchjet{} interval is first scaled by a
value corresponding to the inclusive \pTjet{} cross section, calculated
at Next-to-Leading Order (NLO) with NLL resummation of logarithms in
the jet radius~\cite{Kang_2016}. The 4D response matrix discussed above
is then multiplied by these scaled 2D NLL$^\prime$ predictions (in both
\pTjet{}, ranging from 10 to 200 \GeVc{}, and \ang{})
to obtain the theoretical predictions at charged-hadron level.
To propagate the systematic uncertainty on the original NLL$^\prime$
calculations, this ``folding" procedure is performed individually for
each of fifteen scale variations, from which a total systematic
uncertainty is constructed from the minimum and maximum variation in
each interval. Note that this procedure introduces a model-dependence
to the comparison, and in fact significantly reduces the magnitude
of the systematic uncertainties compared to the parton level; the
repetition of this procedure with both PYTHIA8 and Herwig7 is meant
to estimate the size of this model dependence.

Although the perturbative accuracy of the MC generators is not clear,
by restricting these comparisons to $\pTchjet > 60$ \GeVc{}, there is
adequate matching between the analytical calculations and the
MC generators' final-state parton-level predictions to employ the
nonperturbative corrections via this mapping procedure.
After the folding step, an additional bin-by-bin correction is applied
for multi-parton interactions in the underlying event
using the respective event generator. More specifically, a
ratio is created between the 2D jet angularity distributions
generated with multi-parton interactions on versus off at the
charged-hadron level, which is then multiplied bin-by-bin by the folded
distributions. In all cases, the corrections performed with PYTHIA and
those with Herwig are similar in magnitude, indicating that this correction
procedure is reasonable.

% plots with folding from parton to ch-level
\afterpage{%
\begin{figure}[!t]
\centering{}
\includegraphics[scale=0.8]{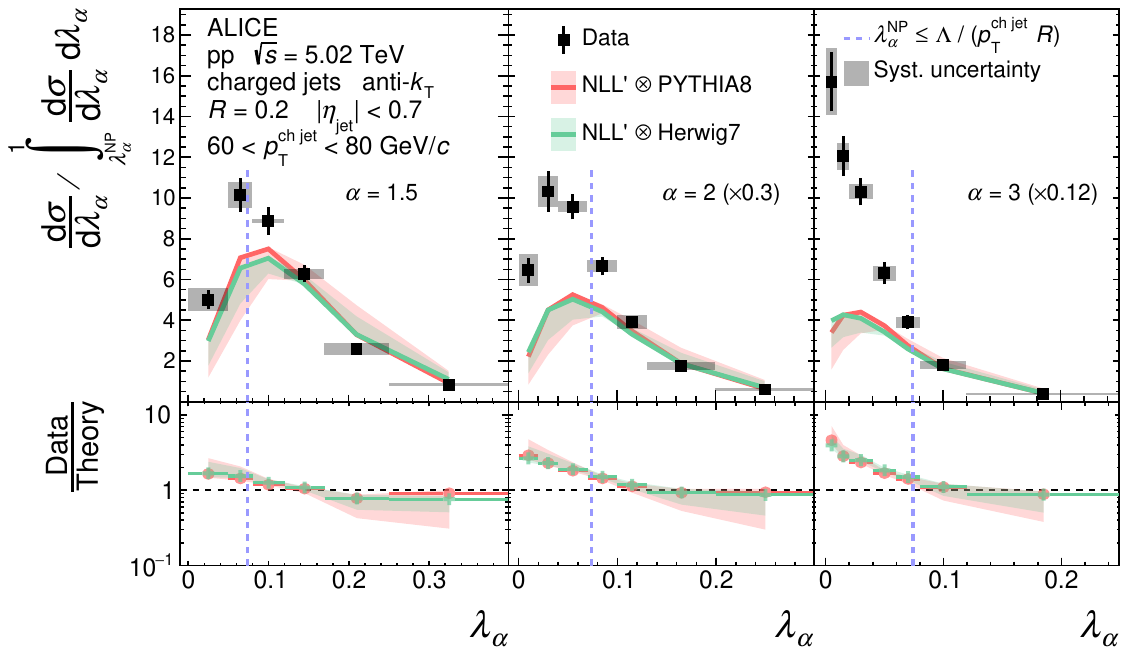}
\includegraphics[scale=0.8]{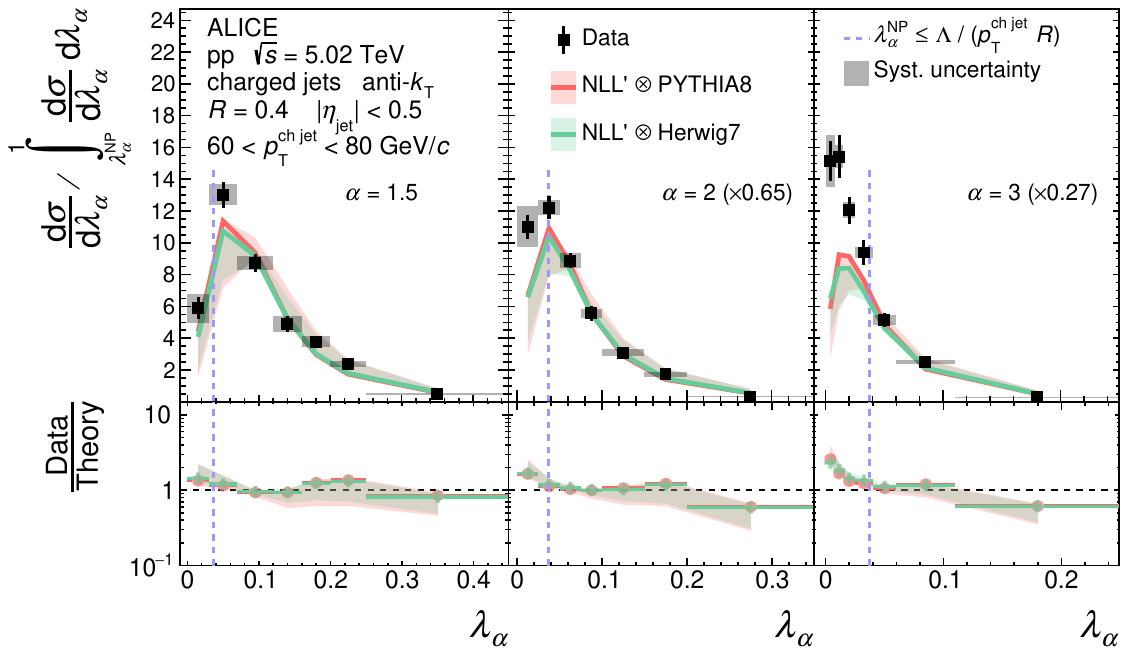}
\caption{Comparison of ungroomed jet angularities \ang{} in \pp{}
collisions for $R=0.2$ (top) and $R=0.4$ (bottom) to analytical
NLL$^\prime$ predictions with MC hadronization corrections in the
range $60 < \pTchjet{} < 80$ \GeVc{}. The distributions are
normalized such that the integral of the perturbative region defined
by $\ang > \angNP{}$ (to the right of the dashed vertical line)
is unity. Divided bins are placed into the left (NP) region.}
\label{fig:theory_ungr_pT60-80}
\end{figure}
\clearpage

% groomed
\begin{figure}[!t]
\centering{}
\includegraphics[scale=0.8]{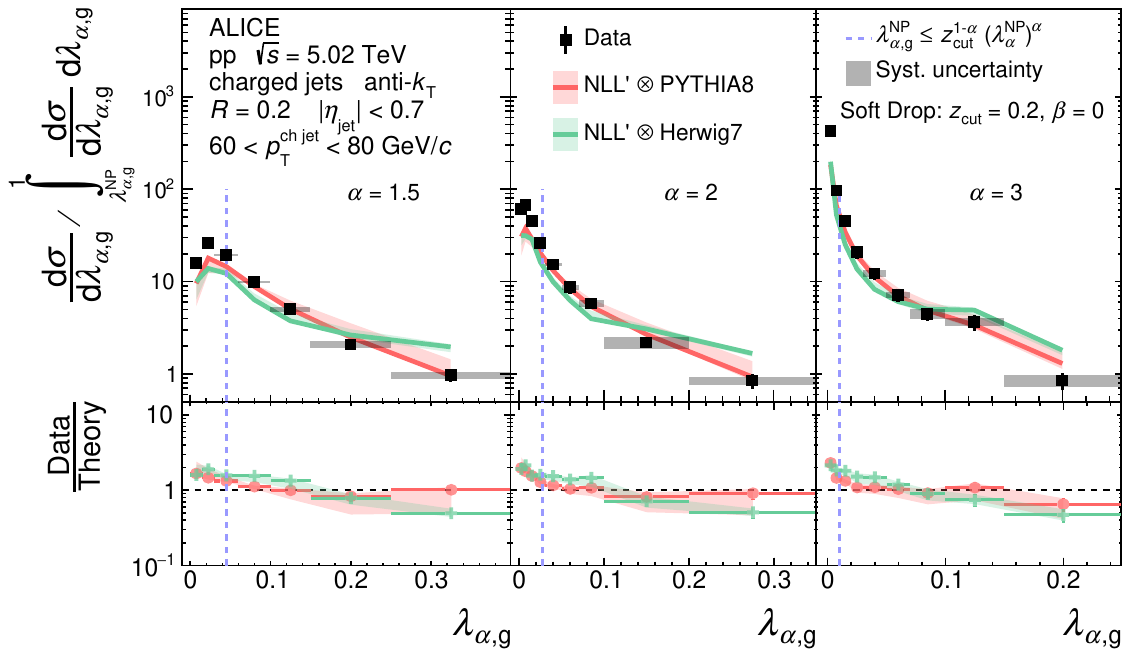}
\includegraphics[scale=0.8]{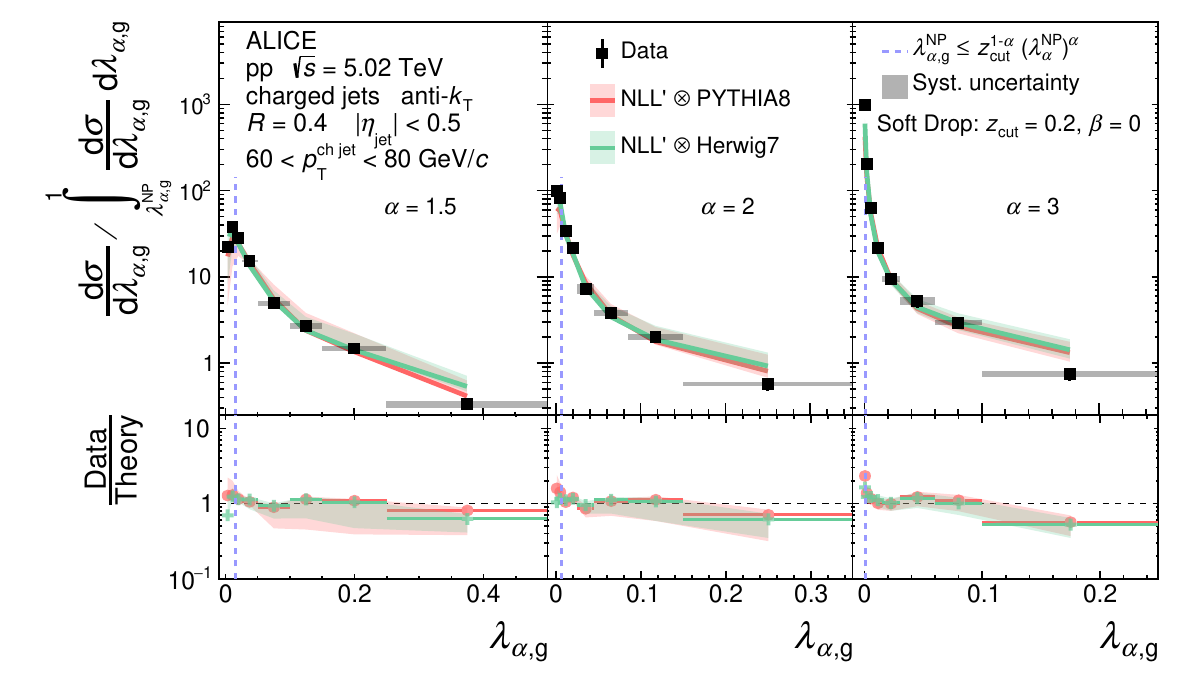}
\caption{Comparison of groomed jet angularities \angsd{} in \pp{}
collisions for $R=0.2$ (top) and $R=0.4$ (bottom) to analytical
NLL$^\prime$ predictions with MC hadronization corrections in the
range $60 < \pTchjet{} < 80$ \GeVc{}. The distributions are
normalized such that the integral of the perturbative region defined
by $\angsd > \angsdNP{}$ (to the right of the dashed vertical line)
is unity. Divided bins are placed into the left (NP) region.}
\label{fig:theory_gr_pT60-80}
\end{figure}
\clearpage
}

Figure~\ref{fig:theory_ungr_pT60-80} shows comparisons of the measured
ungroomed jet angularities to the folded theoretical predictions for
$60 < \pTchjet{} < 80$ \GeVc{},
for both $R=0.2$ (top) and $R=0.4$ (bottom) and for $\alpha = 1.5$ 
(left), 2 (middle), and 3 (right). 
Figure~\ref{fig:theory_gr_pT60-80} shows the corresponding comparisons
for the groomed jet angularities. 
The comparisons for $80 < \pTchjet{} < 100$ \GeVc{} are shown
in Appendix~\ref{appendix-A}.
Predictions for the $\alpha=1$ distributions are not
currently available due to enhanced sensitivity to soft-recoil,
which requires a different factorization~\cite{ang2018}.

A dashed vertical line is drawn as a rough estimate for the division 
of perturbative- and nonperturbative-dominated regions, via
Eq.~\ref{eq:3} or Eq.~\ref{eq:4} with $\Lambda=1$ \GeVc{} and the
mean \pTchjet{} for each interval. Note that the transition from
values of \ang{} which are dominated by perturbative versus
nonperturbative physics is actually smooth, and this vertical line
is merely intended as a visual guide. The nonperturbative-dominated
region of the jet angularities is denoted as \angNP{}.

Since the integral for all of the distributions in Fig.
\ref{fig:ang-1} through Fig.~\ref{fig:angsd-2} is fixed at unity by construction,
it is important to note that disagreement in the
nonperturbative-dominated region induces disagreement in the
perturbative-dominated region. Discrepancy in the nonperturbative
region is expected due to the divergence of $\alpha_s$ and the
corresponding significance of higher-order terms in the perturbative
expansion --- and will necessarily induce disagreement in the perturbative-dominated region.
Accordingly, for these theoretical comparisons, the distributions
are normalized such that the integral above \angNP{} is unity.

%%%%%%%%%%%%%%%%%%%%%%%%%%%%%%%%%%%%%%%%%%%%%%%%
\subsubsection{Shape function based correction}
\label{sect:shape_fn}

An alternate correction technique is also used, which employs a
nonperturbative shape function $F(k)$
\cite{Korchemsky_1999_Fnp, Aschenauer_2020_Fnp, Stewart:2014nna} to correct for
the effects caused by hadronization and the underlying event.
The shape function is defined as
\begin{equation} \label{eqn:F(k)}
F(k) = \frac{4k}{\Omega_\alpha^2}
\exp\left(-\frac{2k}{\Omega_\alpha}\right),
\end{equation}
where $k$ is a momentum scale parameter of the shape function,
and $\Omega_\alpha$ is described by a single parameter
$\Omega=\mathcal{O}(1\;\GeVc)$ obeying the scaling relation
\begin{equation} \label{eqn:omega-scaling}
\Omega_\alpha = \Omega / (\alpha-1),
\end{equation}
and expected to hold universally for hadronization corrections
(but not necessarily for underlying event corrections).
To correct the parton-level calculations
to the hadron level, this shape function is convolved with the
perturbative (parton level) jet angularity distribution via
numerical integration over argument $k$
\begin{equation} \label{eqn:F(k)_convolution}
\frac{\text{d}\sigma}{\text{d}\pTjet{}\text{d}\ang} =
\bigintsss F(k) \frac{\text{d}\sigma_\text{pert}}
                     {\text{d}\pTjet{}\text{d}\ang}
\bigg(\ang - \ang^\text{shift}(k)\bigg)
\; \text{d}k,
\end{equation}
where the shift term $\ang^\text{shift}(k)$ is either
\cite{Aschenauer_2020_Fnp, KANG201941}:
\begin{equation} \label{eqn:F(k)_shift}
\ang^\text{shift}(k) = \frac{k}{\pTjet{} R}
\text{  (ungroomed),} \hspace{1em} \text{or} \hspace{1em}
\zcut^{1-\alpha} \left( \frac{k}{\pTjet{} R} \right)^{\alpha}
\text{  (groomed, with $\beta=0$).}
\end{equation}
The limits of the integral are thus given by the values of $k$ for
which the argument $\left( \ang - \ang^\text{shift}(k) \right)$ 
is between 0 and 1. Since the nonperturbative
parameter $\Omega$ is not calculable within perturbation theory,
four values (0.2, 0.4, 0.8, and 2 \GeVc{}) are chosen to observe the
different shifting effects. These distributions are then corrected
once more using a similar PYTHIA8 folding procedure as described
above to account for the effects of only reconstructing
charged-particle jets. This correction is dominated by a shift and smearing along the \pTjet{} axis.

The comparisons to the ungroomed predictions are shown in Fig.
\ref{fig:theory_ungr_pT60-80_Fnp},
and the groomed predictions are shown in Fig.
\ref{fig:theory_gr_pT60-80_Fnp}.
The shape function approach, specifically the scaling
given in Eq.~\ref{eqn:omega-scaling}, is not fully justified
in the groomed case~\cite{Hoang:2019ceu, Pathak:2020iue}; 
nevertheless, reasonable agreement is observed.
Since this shape convolution does not require matching to MC at
the parton level, the comparisons are extended to the 
$40 < \pTchjet{} < 60$ \GeVc{} interval, but below this the
perturbative accuracy of the parton-level predictions is 
insufficient for rigorous comparisons. 
The comparisons for $40 < \pTchjet{} < 60$ \GeVc{} and 
$80 < \pTchjet{} < 100$ \GeVc{} are shown
in Appendix~\ref{appendix-A}.

% plots with F_NP + folding from hadron to ch-level
\afterpage{%
\begin{figure}[!t]
\centering{}
\includegraphics[scale=0.8]{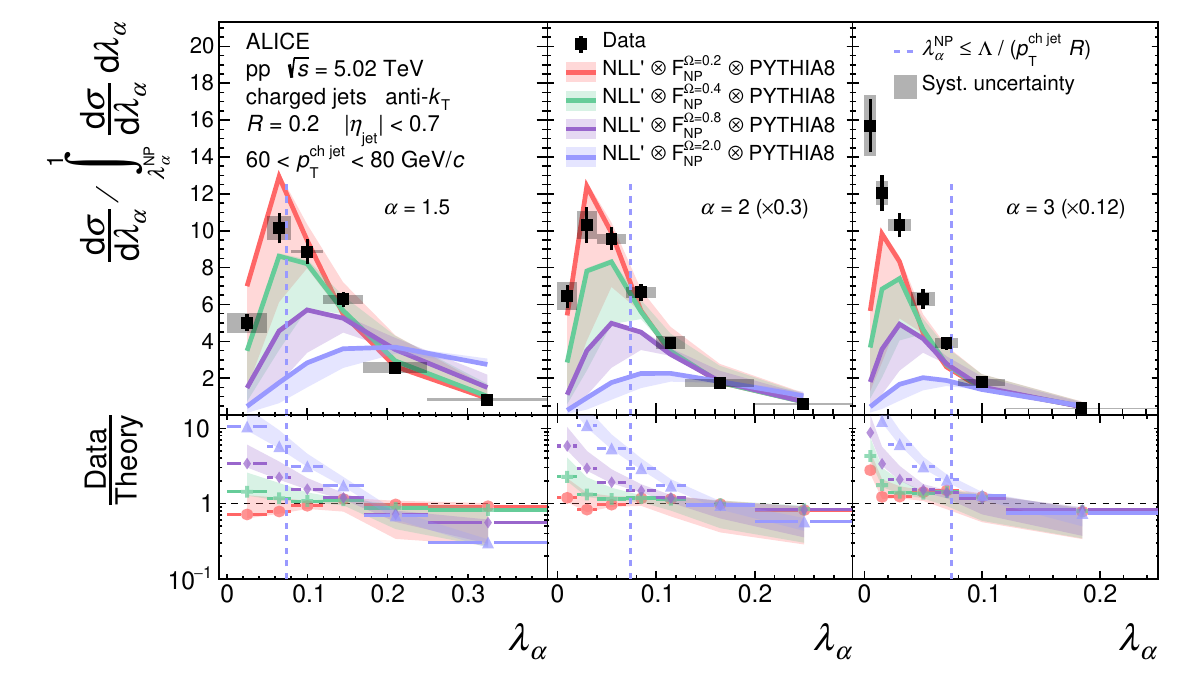}
\includegraphics[scale=0.8]{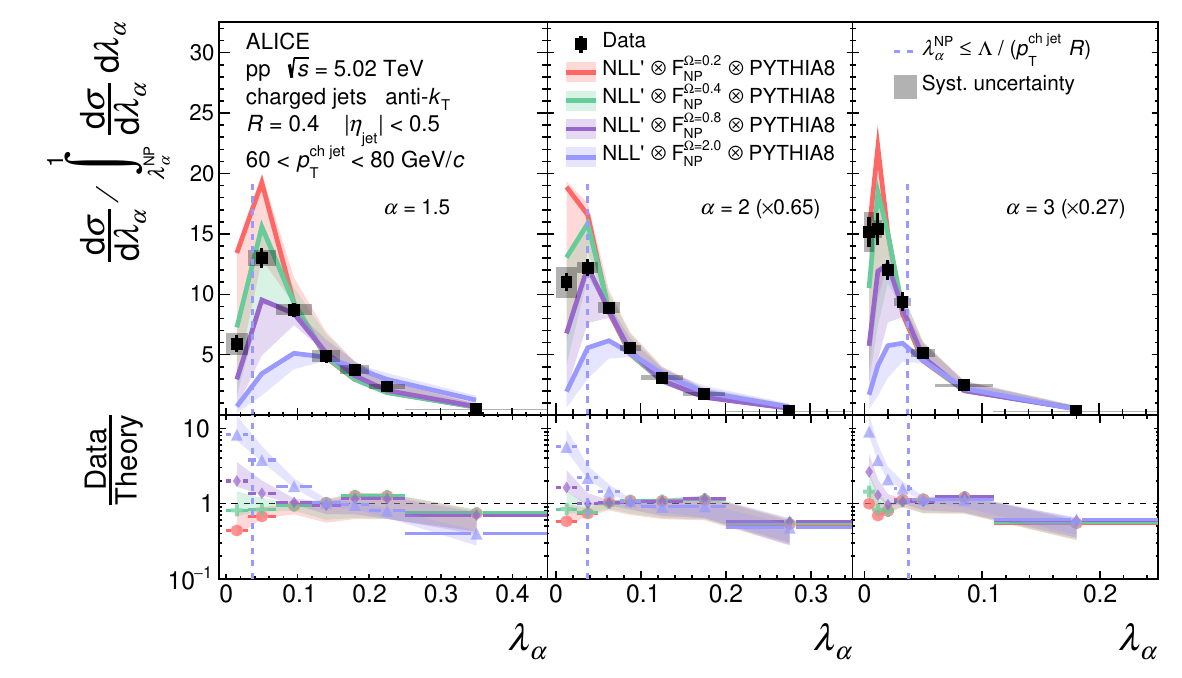}
\caption{Comparison of ungroomed jet angularities \ang{} in \pp{}
collisions for $R=0.2$ (top) and $R=0.4$ (bottom) to analytical
NLL$^\prime$ predictions using $F(k)$ convolution in the
range $60 < \pTchjet{} < 80$ \GeVc{}. The distributions are
normalized such that the integral of the perturbative region defined
by $\ang > \angNP{}$ (to the right of the dashed vertical line)
is unity. Divided bins are placed into the left (NP) region.}
\label{fig:theory_ungr_pT60-80_Fnp}
\end{figure}
\clearpage

% groomed
\begin{figure}[!t]
\centering{}
\includegraphics[scale=0.8]{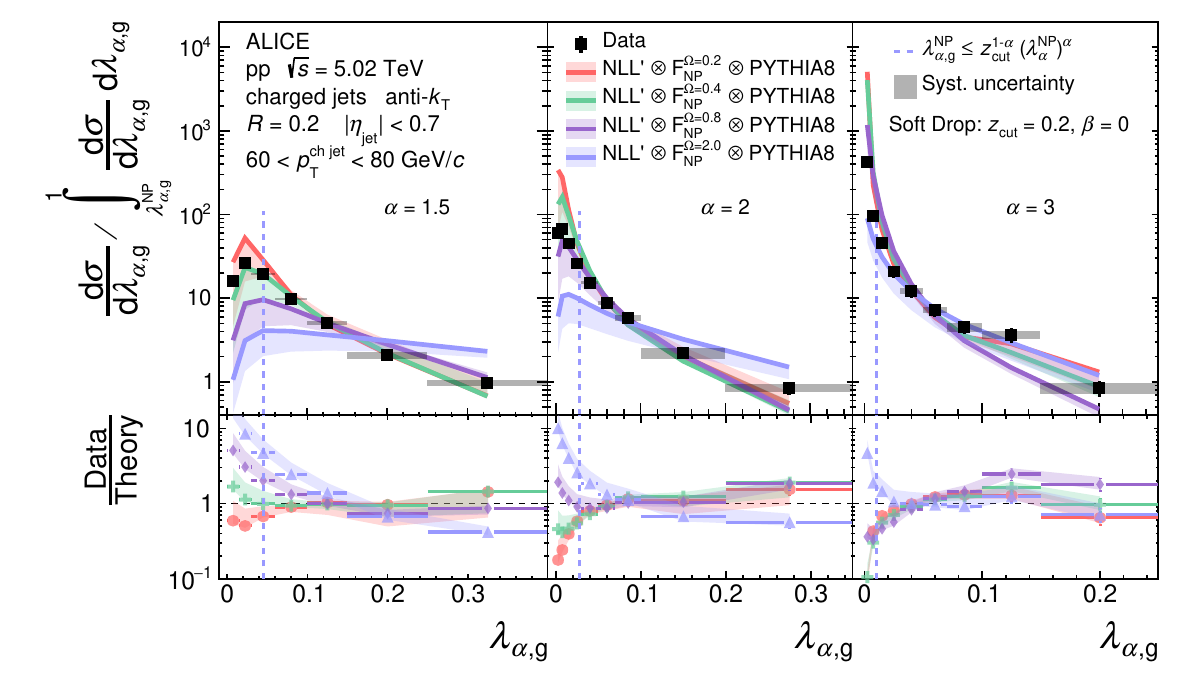}
\includegraphics[scale=0.8]{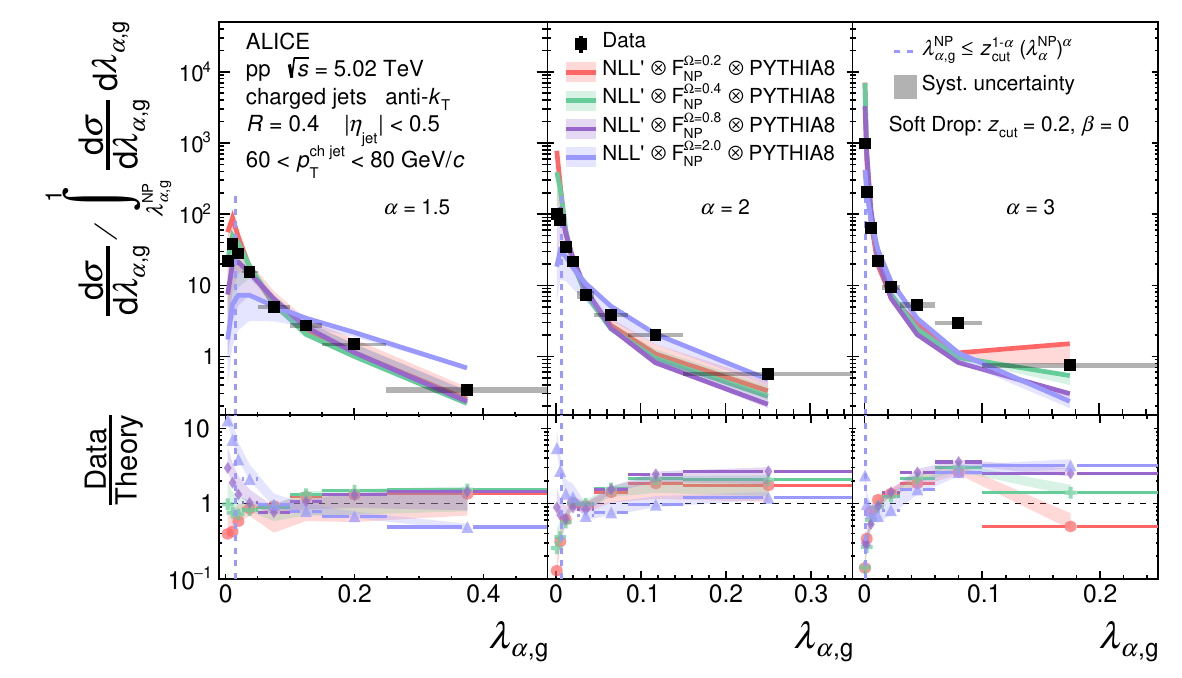}
\caption{Comparison of groomed jet angularities \angsd{} in \pp{}
collisions for $R=0.2$ (top) and $R=0.4$ (bottom) to analytical
NLL$^\prime$ predictions using $F(k)$ convolution in the
range $60 < \pTchjet{} < 80$ \GeVc{}. The distributions are
normalized such that the integral of the perturbative region defined
by $\angsd > \angsdNP{}$ (to the right of the dashed vertical line)
is unity. Divided bins are placed into the left (NP) region.}
\label{fig:theory_gr_pT60-80_Fnp}
\end{figure}
\clearpage
}

%%%%%%%%%%%%%%%%%%%%%%%%%%%%%%%%%%%%%%%%%%%%%%%%%%%%%%%%%%%%%%%%%%
\subsection{Discussion}

% Discuss perturbative regime
The \ang{} distributions are generally consistent with
the calculations within uncertainties when \ang{} is sufficiently
large to be in the pQCD regime. 
This holds approximately independent of $\alpha$, $R$, and $\pTjet{}$, 
and whether or not the jets are groomed.
In some distributions, however, particularly for $R=0.4$, 
modest disagreement is observed at large \ang{}. 
This disagreement cannot be unambiguously associated
with a particular value of \ang{} due to the self-normalization of the 
observable, but rather demonstrates an overall inconsistency
in the shape of the distribution. 
This disagreement could be caused by the unaccounted power corrections in \ang{}, 
or other effects --- and suggests a need for further
theoretical investigation.
Nevertheless, the overall agreement with the perturbative calculations
is striking, given the low-to-moderate jet \pT{} and $R$
considered.

% Discuss nonperturbative regime
For $\alpha=1.5$, the majority of the distributions can
be described perturbatively, as \angNP{} is confined towards the left-hand side of the distributions. As 
$\alpha$ increases to $\alpha=3$, the influence of the \angNP{} region grows, and the ungroomed distributions become strongly nonperturbative.
Similarly, as $R$ increases from $R=0.2$ to $R=0.4$, or as 
\pTchjet{} increases, the size of the perturbative region increases.
In the nonperturbative region $\ang < \angNP$, the \ang{} 
distributions diverge from the calculations.
This is expected, since the perturbative approximations
break down for $\ang < \angNP$, and neither the MC or
shape function corrections are necessarily expected to fully correct
for missing physics at higher orders or for nonperturbative coupling.
In some distributions, the shape-function-based correction
is sometimes able to describe the data partially into the
nonperturbative regime for suitable values of $\Omega$. 

% Discuss groomed vs. ungroomed
While the overall level of agreement is comparable in both the
ungroomed and groomed cases, grooming widens the pQCD regime, as
indicated by the location of the dashed blue line in Figures
[\ref{fig:theory_ungr_pT60-80}-\ref{fig:theory_gr_pT60-80_Fnp}].
On the other hand, grooming shifts the distributions themselves
to significantly smaller values of \ang{}. Nevertheless, this
highlights the potential benefit of grooming in heavy-ion collisions
in order to retain a larger degree of perturbative control in
addition to controlling effects of the underlying event.

% Discuss Omega
The performance of the two nonperturbative correction methods ---
based entirely on MC generators, or on shape functions --- 
are comparable in the perturbative regime. 
Comparing different values of $\Omega$ for the ungroomed case,
where Eq.~\ref{eqn:omega-scaling} is justified, there is 
in many cases only a small difference between the 
calculations with $\Omega=0.2$, 0.4, and 0.8 \GeVc{}. 
However, for $\alpha=1.5$ and $\alpha=2$, larger values
of $\Omega$ ($\Omega=2$ \GeVc{}) appear to have more tension with
the data in the perturbative regime than smaller values.
For $\alpha=3$, the perturbative region is too small to make any clear statement.
One must bear in mind, however, that \angNP\ is only a rough
characterization of the regime of validity of the perturbative calculation.
Consequently, it is unknown whether this disagreement is due to the value of $\Omega$ or due to the breakdown of the perturbative calculation.
For smaller values of $\Omega$ (e.g. $\Omega=0.2$ or $0.4$ \GeVc{}),
the predicted scaling of Eq.~\ref{eqn:omega-scaling} is
consistent with the data.
Note that the value of $\Omega$ which describes the data
is $\mathcal{O}(1)$ as expected for hadronization corrections. 
These smaller values contrast with a previous
result of $\Omega=3.5$ \GeVc{} for the ungroomed mass of $R=0.4$ jets at $200 < \pTjet{} < 300$ \GeVc{}~\cite{Kang:2018jwa},
suggesting that the underlying event contribution to $\Omega$, which is not expected to obey the scaling of Eq.~\ref{eqn:omega-scaling}, may be modified by jets measured at different \pTjet{} or by the choice to reconstruct jets using only charged-particle tracks.
No significant $R$-dependence is observed in the scaling behavior
in this analysis, suggesting that any scaling-breaking 
underlying event contributions, when also combined with hadronization corrections, are small for $R=0.2$ and 0.4.

%% file: Ch6_Conclusion.tex
\section{Conclusion}
The generalized jet angularities are reported both with and
without Soft Drop grooming, \angsd{} and \ang{}, respectively,
for charged-particle jets in \pp{} collisions
at $\sqrt{s}=5.02$ TeV with the ALICE detector.
This measurement of both the ungroomed and, for the first time, the
groomed jet angularities provides constraints on models and captures
the interplay between perturbative and nonperturbative effects in QCD.
Systematic variations of the contributions from collinear and soft
radiation of the shower, captured within a given $R$, are provided by
measuring the jet angularities for a selection of $\alpha$ parameters.
These results consequently provide rigorous tests of pQCD calculations.

The theoretical predictions at NLL$^\prime$ in SCET show an overall
agreement with the data for jets with values of $\ang$ in the
perturbative regime delimited by a collinear-soft momentum scale in the factorization framework of about 1$~\GeVc$.
The calculations, after accounting for nonperturbative effects
by two different methods, are compatible within about 20\% or 
better with the data in the perturbative region for all explored
values of $R$ and $\alpha$. However, larger deviations of up to
about 50\% are observed in the tails of some distributions,
suggesting a need for further theoretical study. By making comparisons
solely in the perburbatively-dominated regime, consistency is seen
with a predicted universal scaling of the nonperturbative shape
function parameter $\Omega_\alpha$ with value $\Omega<1$.
A clear breakdown of the agreement is observed for small $\lambda$,
where the perturbative calculation is expected to fail.
Such nonperturbative effects include soft splittings and hadronization,
and these effects dominate over significant regions of the phase space
of moderate and low-energy jets.
This is corroborated by the comparison of the measured groomed jet
angularities to the equivalent theoretical calculations,
which demonstrate a wider range of agreement with the perturbative
calculations.

These comparisons provide critical guidance for measurements in
high-energy heavy-ion collisions where the internal structure of jets
may undergo modifications via scatterings of jet fragments with the
hot and dense QCD medium. Our measurements demonstrate that any
comparison to pQCD must also consider the regimes of \ang{} and
\angsd{} that are controlled by perturbative processes as opposed to
those that are dominated by nonperturbative processes. This provides
guidance for the selections of $\alpha$, $R$, and $\pTchjet$, and
indicates the importance of capturing the complete spectrum of
processes (perturbative and non-perturbative) in theory calculations
attempting to explain jet quenching.
 
These measurements further highlight that disagreement between
theoretical predictions and data in the nonperturbative regime will
necessarily induce disagreement in the perturbative regime, when in
fact the perturbative accuracy of predictions should only be
scrutinized within the perturbative regime.
In practice, these measurements give a clear indication that careful
inspection is needed when interpreting measurements of jet
substructure based on models of jet quenching in heavy-ion collisions
for observables including the jet angularity and the jet mass.
Future measurements will benefit from the provided
guidance demonstrating not only the agreement of jet angularities
with pQCD calculations in the perturbative regime but also 
on selecting on jet angularity differentially with $\alpha$, $R$, and
$\pTchjet$ in order to maximize theoretical control and interpretation
of the perturbative and nonpertubative regimes of jet substructure
observables.

%% file: fa_2021-07-05.tex
% Version: 2021-07-05

The ALICE Collaboration would like to thank all its engineers and technicians for their invaluable contributions to the construction of the experiment and the CERN accelerator teams for the outstanding performance of the LHC complex.
The ALICE Collaboration gratefully acknowledges the resources and support provided by all Grid centres and the Worldwide LHC Computing Grid (WLCG) collaboration.
The ALICE Collaboration acknowledges the following funding agencies for their support in building and running the ALICE detector:
A. I. Alikhanyan National Science Laboratory (Yerevan Physics Institute) Foundation (ANSL), State Committee of Science and World Federation of Scientists (WFS), Armenia;
Austrian Academy of Sciences, Austrian Science Fund (FWF): [M 2467-N36] and Nationalstiftung f\"{u}r Forschung, Technologie und Entwicklung, Austria;
Ministry of Communications and High Technologies, National Nuclear Research Center, Azerbaijan;
Conselho Nacional de Desenvolvimento Cient\'{\i}fico e Tecnol\'{o}gico (CNPq), Financiadora de Estudos e Projetos (Finep), Funda\c{c}\~{a}o de Amparo \`{a} Pesquisa do Estado de S\~{a}o Paulo (FAPESP) and Universidade Federal do Rio Grande do Sul (UFRGS), Brazil;
Ministry of Education of China (MOEC) , Ministry of Science \& Technology of China (MSTC) and National Natural Science Foundation of China (NSFC), China;
Ministry of Science and Education and Croatian Science Foundation, Croatia;
Centro de Aplicaciones Tecnol\'{o}gicas y Desarrollo Nuclear (CEADEN), Cubaenerg\'{\i}a, Cuba;
Ministry of Education, Youth and Sports of the Czech Republic, Czech Republic;
The Danish Council for Independent Research | Natural Sciences, the VILLUM FONDEN and Danish National Research Foundation (DNRF), Denmark;
Helsinki Institute of Physics (HIP), Finland;
Commissariat \`{a} l'Energie Atomique (CEA) and Institut National de Physique Nucl\'{e}aire et de Physique des Particules (IN2P3) and Centre National de la Recherche Scientifique (CNRS), France;
Bundesministerium f\"{u}r Bildung und Forschung (BMBF) and GSI Helmholtzzentrum f\"{u}r Schwerionenforschung GmbH, Germany;
General Secretariat for Research and Technology, Ministry of Education, Research and Religions, Greece;
National Research, Development and Innovation Office, Hungary;
Department of Atomic Energy Government of India (DAE), Department of Science and Technology, Government of India (DST), University Grants Commission, Government of India (UGC) and Council of Scientific and Industrial Research (CSIR), India;
Indonesian Institute of Science, Indonesia;
Istituto Nazionale di Fisica Nucleare (INFN), Italy;
Institute for Innovative Science and Technology , Nagasaki Institute of Applied Science (IIST), Japanese Ministry of Education, Culture, Sports, Science and Technology (MEXT) and Japan Society for the Promotion of Science (JSPS) KAKENHI, Japan;
Consejo Nacional de Ciencia (CONACYT) y Tecnolog\'{i}a, through Fondo de Cooperaci\'{o}n Internacional en Ciencia y Tecnolog\'{i}a (FONCICYT) and Direcci\'{o}n General de Asuntos del Personal Academico (DGAPA), Mexico;
Nederlandse Organisatie voor Wetenschappelijk Onderzoek (NWO), Netherlands;
The Research Council of Norway, Norway;
Commission on Science and Technology for Sustainable Development in the South (COMSATS), Pakistan;
Pontificia Universidad Cat\'{o}lica del Per\'{u}, Peru;
Ministry of Education and Science, National Science Centre and WUT ID-UB, Poland;
Korea Institute of Science and Technology Information and National Research Foundation of Korea (NRF), Republic of Korea;
Ministry of Education and Scientific Research, Institute of Atomic Physics and Ministry of Research and Innovation and Institute of Atomic Physics, Romania;
Joint Institute for Nuclear Research (JINR), Ministry of Education and Science of the Russian Federation, National Research Centre Kurchatov Institute, Russian Science Foundation and Russian Foundation for Basic Research, Russia;
Ministry of Education, Science, Research and Sport of the Slovak Republic, Slovakia;
National Research Foundation of South Africa, South Africa;
Swedish Research Council (VR) and Knut \& Alice Wallenberg Foundation (KAW), Sweden;
European Organization for Nuclear Research, Switzerland;
Suranaree University of Technology (SUT), National Science and Technology Development Agency (NSDTA) and Office of the Higher Education Commission under NRU project of Thailand, Thailand;
Turkish Energy, Nuclear and Mineral Research Agency (TENMAK), Turkey;
National Academy of  Sciences of Ukraine, Ukraine;
Science and Technology Facilities Council (STFC), United Kingdom;
National Science Foundation of the United States of America (NSF) and United States Department of Energy, Office of Nuclear Physics (DOE NP), United States of America.

%% file: AppendixA.tex
\section{Additional figures} \label{appendix-A}

\begin{figure}[!b]
\centering{}
\includegraphics[scale=0.8]{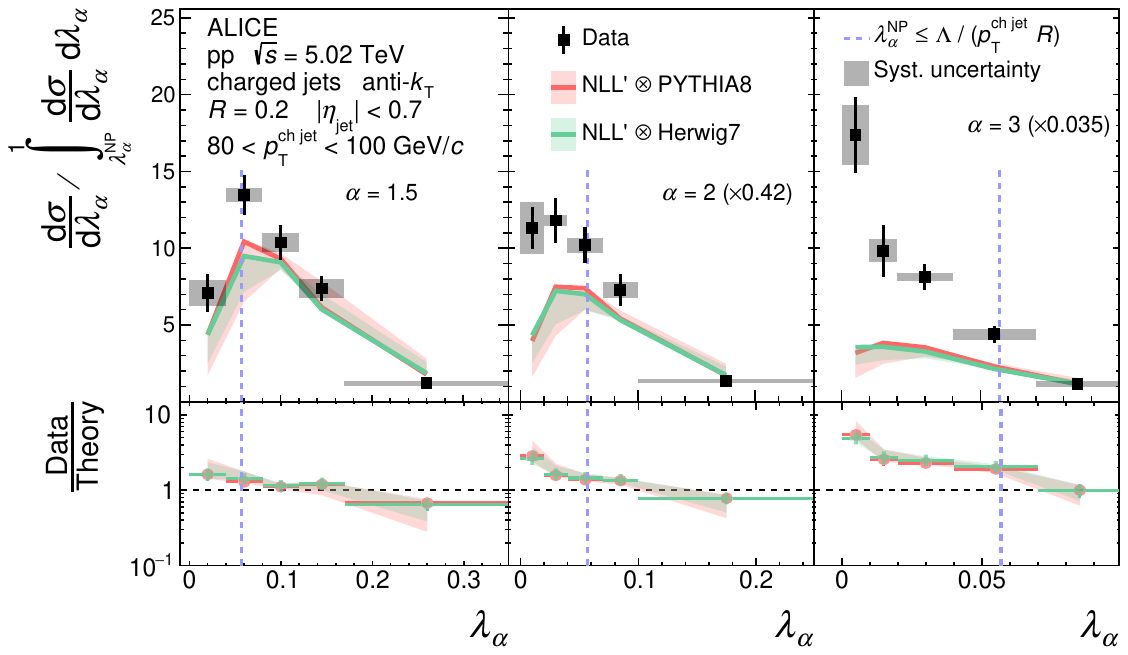}
\includegraphics[scale=0.8]{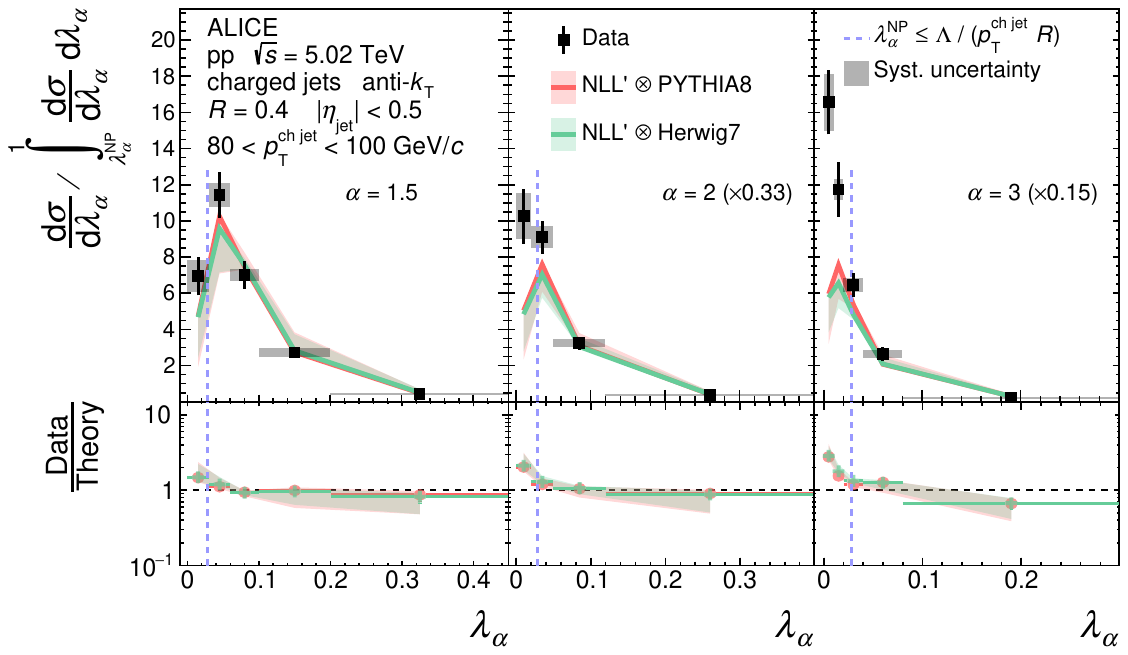}
\caption{Comparison of ungroomed jet angularities \ang{} in \pp{}
collisions for $R=0.2$ (top) and $R=0.4$ (bottom) to analytical
NLL$^\prime$ predictions with MC hadronization corrections in the
range $80 < \pTchjet{} < 100$ \GeVc{}. The distributions are
normalized such that the integral of the perturbative region defined
by $\ang > \angNP{}$ (to the right of the dashed vertical line)
is unity. Divided bins are placed into the left (NP) region.}
\label{fig:theory_ungr_pT80-100}
\end{figure}

\begin{figure}[!t]
\centering{}
\includegraphics[scale=0.8]{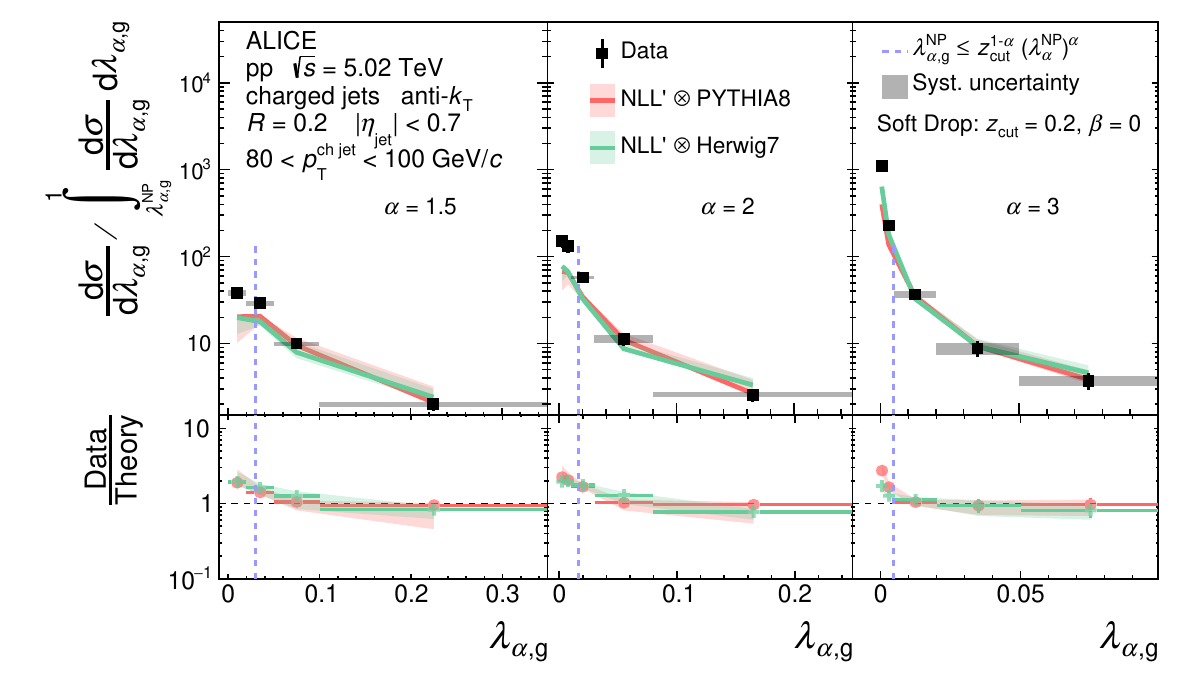}
\includegraphics[scale=0.8]{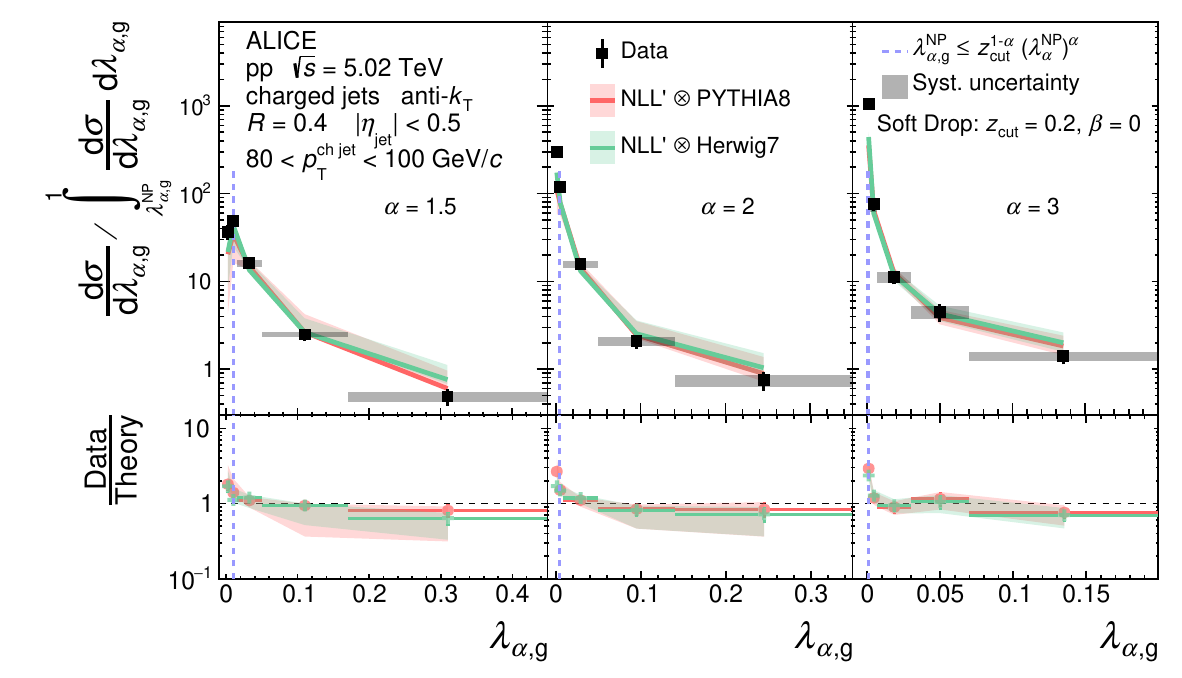}
\caption{Comparison of groomed jet angularities \angsd{} in \pp{}
collisions for $R=0.2$ (top) and $R=0.4$ (bottom) to analytical
NLL$^\prime$ predictions with MC hadronization corrections in the
range $80 < \pTchjet{} < 100$ \GeVc{}. The distributions are
normalized such that the integral of the perturbative region defined
by $\angsd > \angsdNP{}$ (to the right of the dashed vertical line)
is unity. Divided bins are placed into the left (NP) region.}
\label{fig:theory_gr_pT80-100}
\end{figure}

Figures~\ref{fig:theory_ungr_pT80-100},~\ref{fig:theory_gr_pT80-100}
show the groomed and ungroomed angularities for
$80 < \pTchjet{} < 100$ \GeVc{} using MC generators to
apply hadronization corrections.

\begin{figure}[!t]
\centering{}
\includegraphics[scale=0.8]{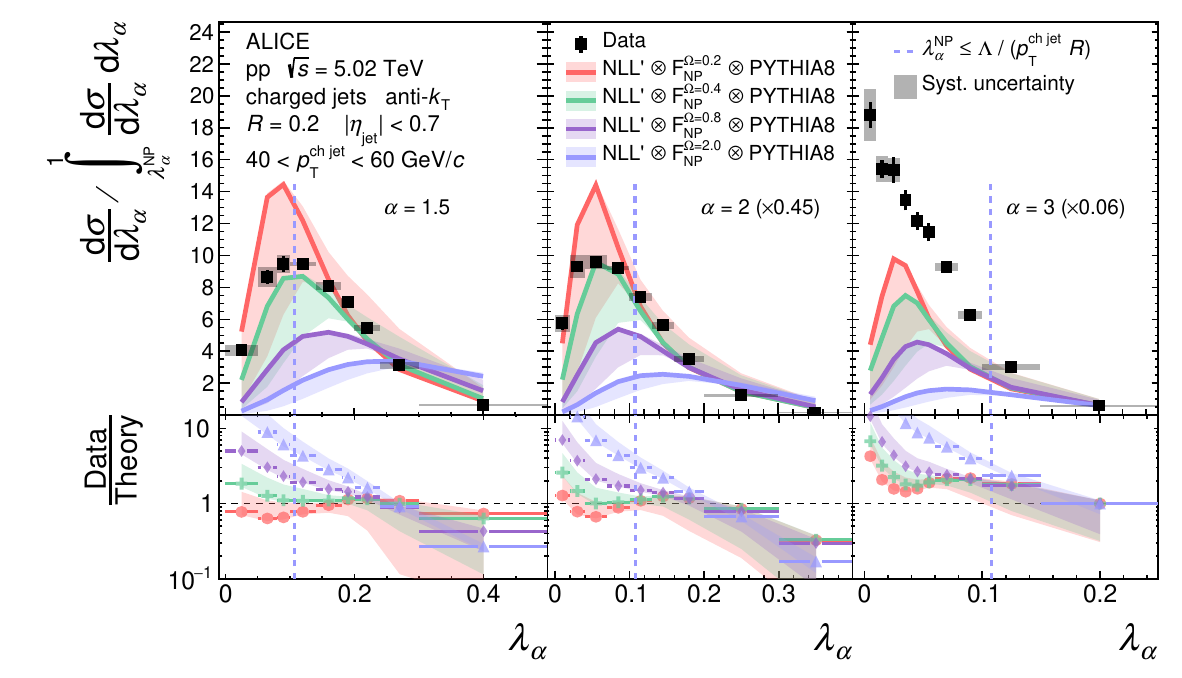}
\includegraphics[scale=0.8]{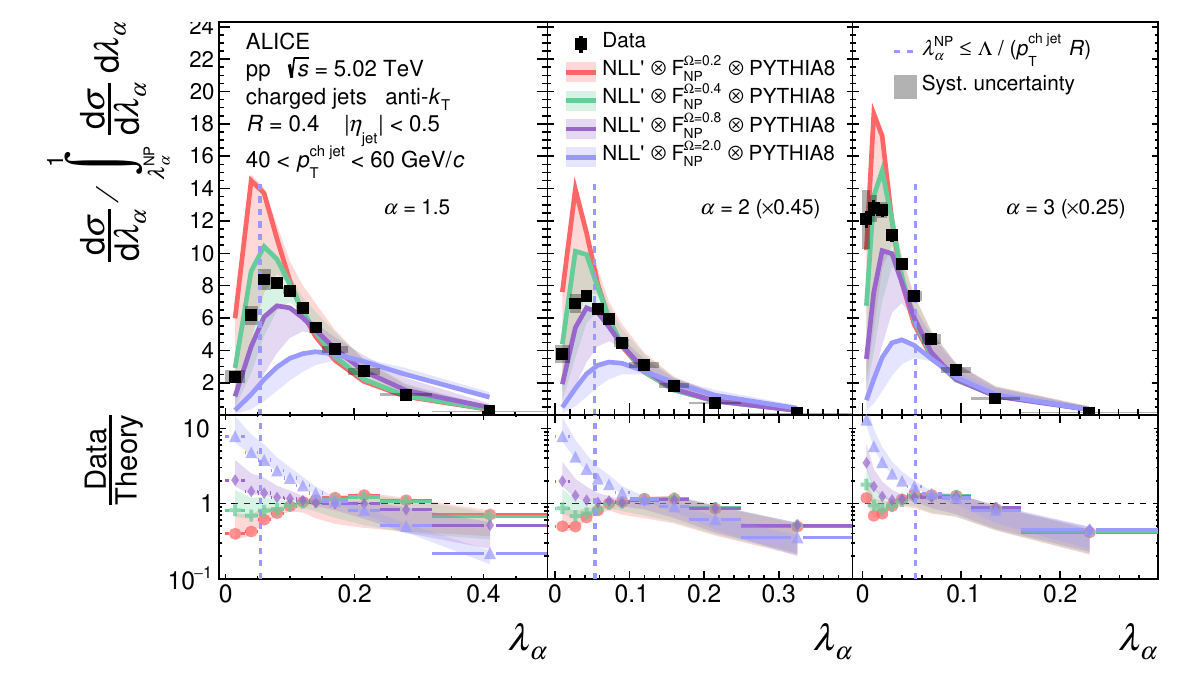}
\caption{Comparison of ungroomed jet angularities \ang{} in \pp{}
collisions for $R=0.2$ (top) and $R=0.4$ (bottom) to analytical
NLL$^\prime$ predictions using $F(k)$ convolution in the
range $40 < \pTchjet{} < 60$ \GeVc{}. The distributions are
normalized such that the integral of the perturbative region defined
by $\ang > \angNP{}$ (to the right of the dashed vertical line)
is unity. Divided bins are placed into the left (NP) region.}
\label{fig:theory_ungr_pT40-60_Fnp}
\end{figure}

\begin{figure}[!t]
\centering{}
\includegraphics[scale=0.8]{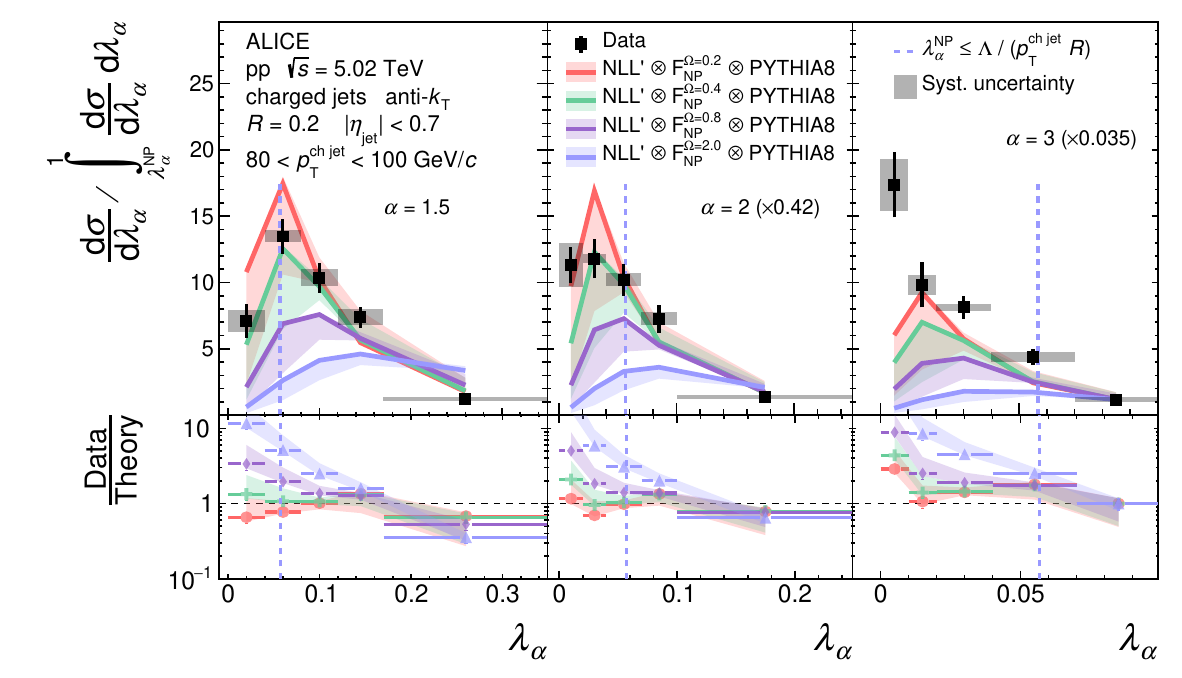}
\includegraphics[scale=0.8]{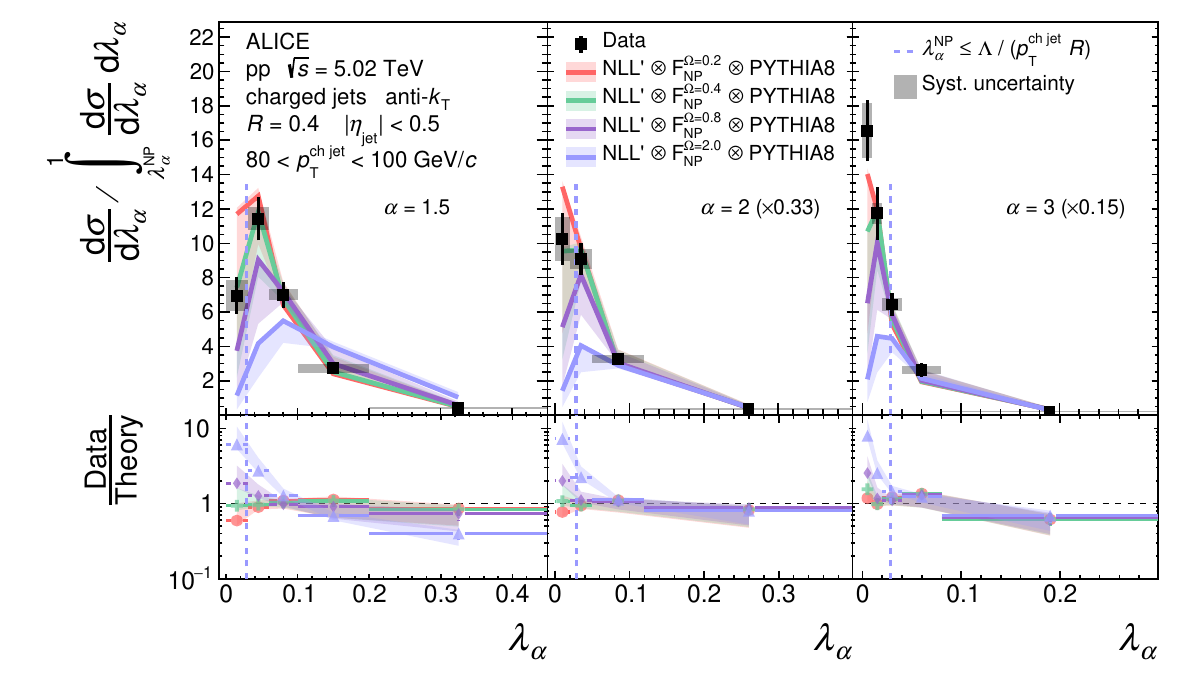}
\caption{Comparison of ungroomed jet angularities \ang{} in \pp{}
collisions for $R=0.2$ (top) and $R=0.4$ (bottom) to analytical
NLL$^\prime$ predictions using $F(k)$ convolution in the
range $80 < \pTchjet{} < 100$ \GeVc{}. The distributions are
normalized such that the integral of the perturbative region defined
by $\ang > \angNP{}$ (to the right of the dashed vertical line)
is unity. Divided bins are placed into the left (NP) region.}
\label{fig:theory_ungr_pT80-100_Fnp}
\end{figure}

\begin{figure}[!t]
\centering{}
\includegraphics[scale=0.8]{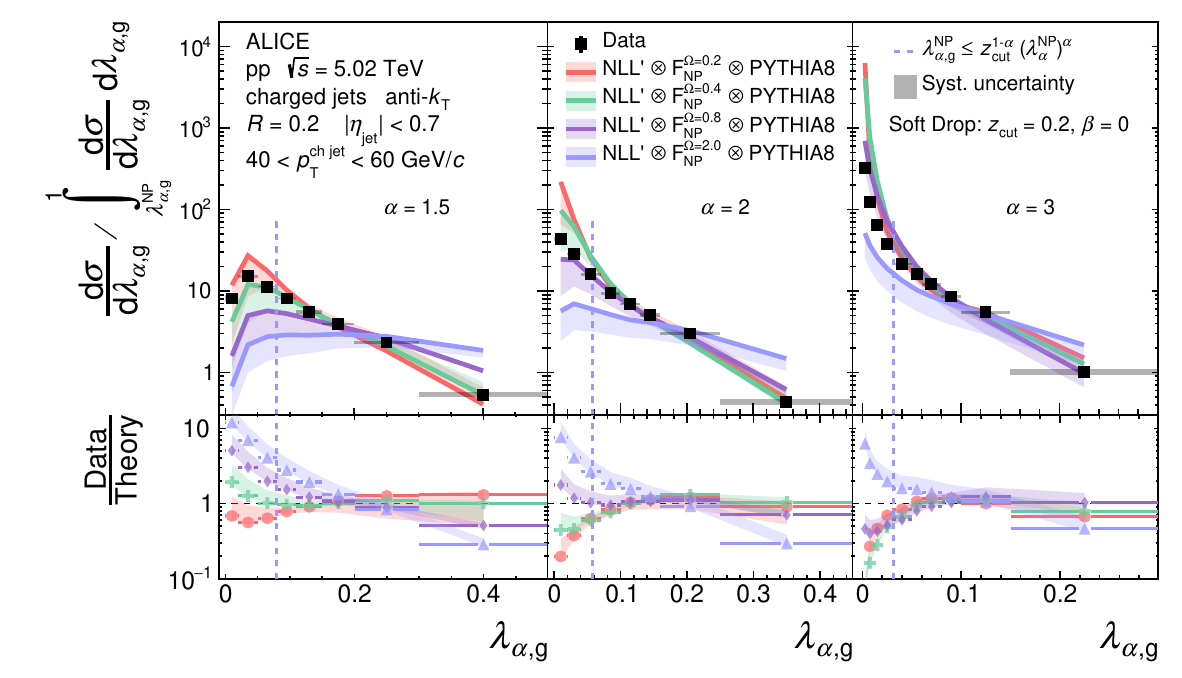}
\includegraphics[scale=0.8]{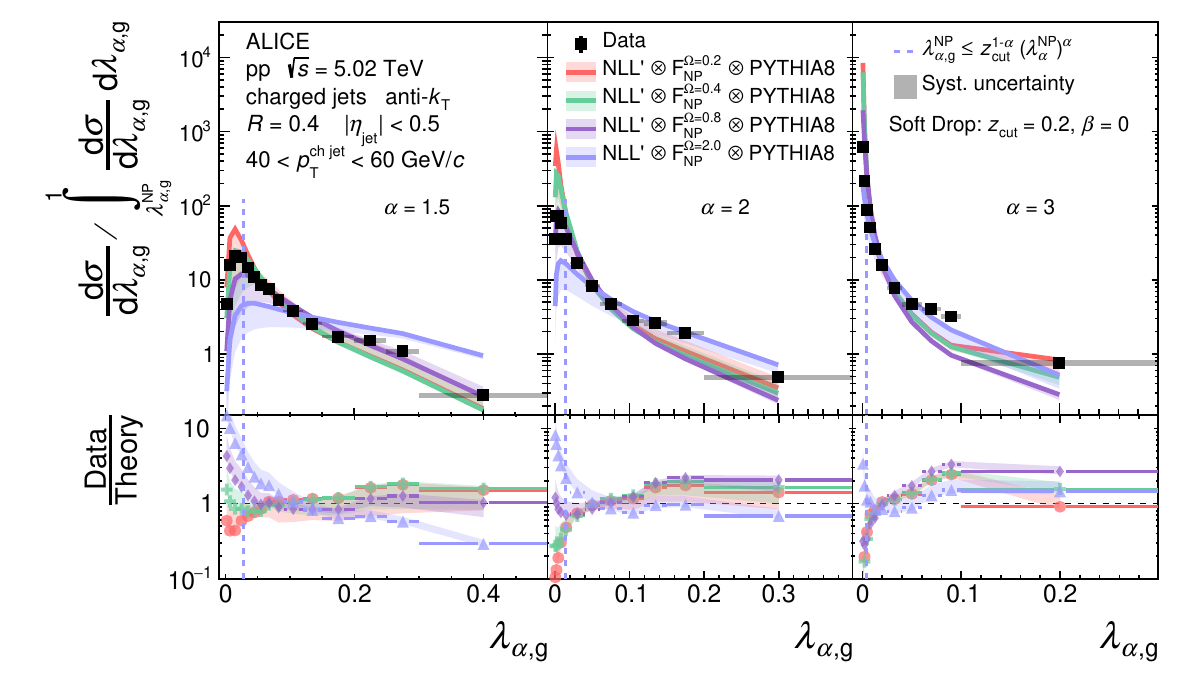}
\caption{Comparison of groomed jet angularities \angsd{} in \pp{}
collisions for $R=0.2$ (top) and $R=0.4$ (bottom) to analytical
NLL$^\prime$ predictions using $F(k)$ convolution in the
range $40 < \pTchjet{} < 60$ \GeVc{}. The distributions are
normalized such that the integral of the perturbative region defined
by $\angsd > \angsdNP{}$ (to the right of the dashed vertical line)
is unity. Divided bins are placed into the left (NP) region.}
\label{fig:theory_gr_pT40-60_Fnp}
\end{figure}

\begin{figure}[!t]
\centering{}
\includegraphics[scale=0.8]{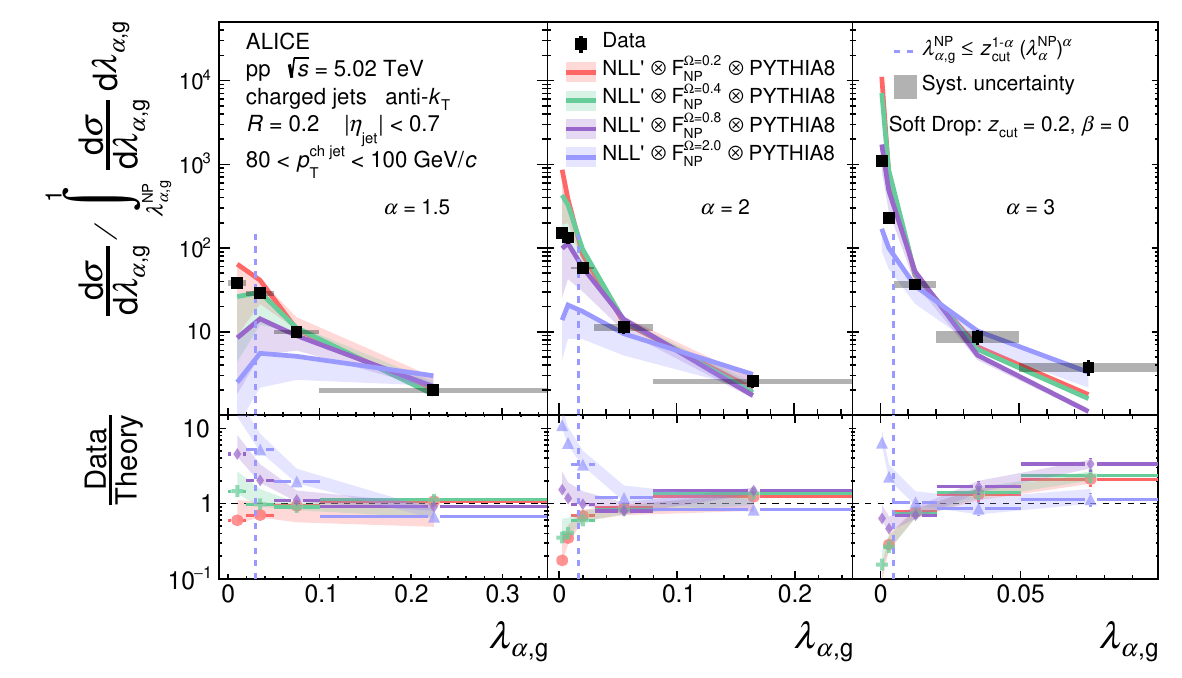}
\includegraphics[scale=0.8]{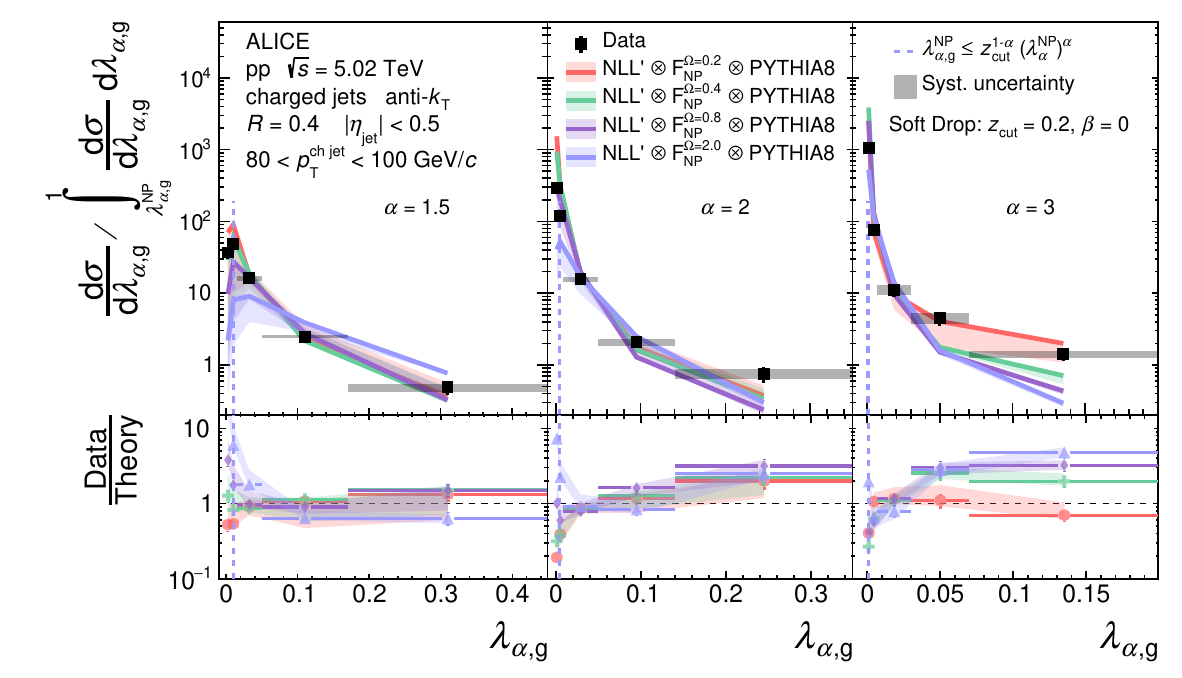}
\caption{Comparison of groomed jet angularities \angsd{} in \pp{}
collisions for $R=0.2$ (top) and $R=0.4$ (bottom) to analytical
NLL$^\prime$ predictions using $F(k)$ convolution in the
range $80 < \pTchjet{} < 100$ \GeVc{}. The distributions are
normalized such that the integral of the perturbative region defined
by $\angsd > \angsdNP{}$ (to the right of the dashed vertical line)
is unity. Divided bins are placed into the left (NP) region.}
\label{fig:theory_gr_pT80-100_Fnp}
\end{figure}

Figures~\ref{fig:theory_ungr_pT40-60_Fnp},~\ref{fig:theory_ungr_pT80-100_Fnp} and 
\ref{fig:theory_gr_pT40-60_Fnp},~\ref{fig:theory_gr_pT80-100_Fnp}
show the groomed and ungroomed angularities using the shape function
to apply hadronization corrections, for
$40 < \pTchjet{} < 60$ \GeVc{} and $80 < \pTchjet{} < 100$ \GeVc{}.

% PYTHIA/HERWIG
\clearpage

\clearpage

% Shape function
\clearpage

\clearpage

\clearpage

\clearpage

%% file: 2021-07-05-Alice_Authorlist_2021-07-05.tex
% ALICE Collaboration author list for 2021-07-05
\small
\begin{flushleft}

S.~Acharya$^{\rm 143}$, 
D.~Adamov\'{a}$^{\rm 98}$, 
A.~Adler$^{\rm 76}$, 
G.~Aglieri Rinella$^{\rm 35}$, 
M.~Agnello$^{\rm 31}$, 
N.~Agrawal$^{\rm 55}$, 
Z.~Ahammed$^{\rm 143}$, 
S.~Ahmad$^{\rm 16}$, 
S.U.~Ahn$^{\rm 78}$, 
I.~Ahuja$^{\rm 39}$
% ...

\section*{Affiliation notes}

$^{\rm I}$ Deceased\\
$^{\rm II}$ Also at: Italian National Agency for New Technologies, Energy and Sustainable Economic Development (ENEA), Bologna, Italy\\
$^{\rm III}$ Also at: Dipartimento DET del Politecnico di Torino, Turin, Italy\\
$^{\rm IV}$ Also at: M.V. Lomonosov Moscow State University, D.V. Skobeltsyn Institute of Nuclear, Physics, Moscow, Russia\\
$^{\rm V}$ Also at: Department of Applied Physics, Aligarh Muslim University, Aligarh, India
\\
$^{\rm VI}$ Also at: Institute of Theoretical Physics, University of Wroclaw, Poland\\

\section*{Collaboration Institutes}

$^{1}$ A.I. Alikhanyan National Science Laboratory (Yerevan Physics Institute) Foundation, Yerevan, Armenia\\
$^{2}$ AGH University of Science and Technology, Cracow, Poland\\
$^{3}$ Bogolyubov Institute for Theoretical Physics, National Academy of Sciences of Ukraine, Kiev, Ukraine\\
$^{4}$ Bose Institute, Department of Physics  and Centre for Astroparticle Physics and Space Science (CAPSS), Kolkata, India\\
$^{5}$ Budker Institute for Nuclear Physics, Novosibirsk, Russia\\
$^{6}$ California Polytechnic State University, San Luis Obispo, California, United States\\
$^{7}$ Central China Normal University, Wuhan, China\\
$^{8}$ Centro de Aplicaciones Tecnol\'{o}gicas y Desarrollo Nuclear (CEADEN), Havana, Cuba\\
$^{9}$ Centro de Investigaci\'{o}n y de Estudios Avanzados (CINVESTAV), Mexico City and M\'{e}rida, Mexico\\
$^{10}$ Chicago State University, Chicago, Illinois, United States\\
$^{11}$ China Institute of Atomic Energy, Beijing, China\\
$^{12}$ Chungbuk National University, Cheongju, Republic of Korea\\
$^{13}$ Comenius University Bratislava, Faculty of Mathematics, Physics and Informatics, Bratislava, Slovakia\\
$^{14}$ COMSATS University Islamabad, Islamabad, Pakistan\\
$^{15}$ Creighton University, Omaha, Nebraska, United States\\
$^{16}$ Department of Physics, Aligarh Muslim University, Aligarh, India\\
$^{17}$ Department of Physics, Pusan National University, Pusan, Republic of Korea\\
$^{18}$ Department of Physics, Sejong University, Seoul, Republic of Korea\\
$^{19}$ Department of Physics, University of California, Berkeley, California, United States\\
$^{20}$ Department of Physics, University of Oslo, Oslo, Norway\\
$^{21}$ Department of Physics and Technology, University of Bergen, Bergen, Norway\\
$^{22}$ Dipartimento di Fisica dell'Universit\`{a} 'La Sapienza' and Sezione INFN, Rome, Italy\\
$^{23}$ Dipartimento di Fisica dell'Universit\`{a} and Sezione INFN, Cagliari, Italy\\
$^{24}$ Dipartimento di Fisica dell'Universit\`{a} and Sezione INFN, Trieste, Italy\\
$^{25}$ Dipartimento di Fisica dell'Universit\`{a} and Sezione INFN, Turin, Italy\\
$^{26}$ Dipartimento di Fisica e Astronomia dell'Universit\`{a} and Sezione INFN, Bologna, Italy\\
$^{27}$ Dipartimento di Fisica e Astronomia dell'Universit\`{a} and Sezione INFN, Catania, Italy\\
$^{28}$ Dipartimento di Fisica e Astronomia dell'Universit\`{a} and Sezione INFN, Padova, Italy\\
$^{29}$ Dipartimento di Fisica e Nucleare e Teorica, Universit\`{a} di Pavia, Pavia, Italy\\
$^{30}$ Dipartimento di Fisica `E.R.~Caianiello' dell'Universit\`{a} and Gruppo Collegato INFN, Salerno, Italy\\
$^{31}$ Dipartimento DISAT del Politecnico and Sezione INFN, Turin, Italy\\
$^{32}$ Dipartimento di Scienze e Innovazione Tecnologica dell'Universit\`{a} del Piemonte Orientale and INFN Sezione di Torino, Alessandria, Italy\\
$^{33}$ Dipartimento di Scienze MIFT, Universit\`{a} di Messina, Messina, Italy\\
$^{34}$ Dipartimento Interateneo di Fisica `M.~Merlin' and Sezione INFN, Bari, Italy\\
$^{35}$ European Organization for Nuclear Research (CERN), Geneva, Switzerland\\
$^{36}$ Faculty of Electrical Engineering, Mechanical Engineering and Naval Architecture, University of Split, Split, Croatia\\
$^{37}$ Faculty of Engineering and Science, Western Norway University of Applied Sciences, Bergen, Norway\\
$^{38}$ Faculty of Nuclear Sciences and Physical Engineering, Czech Technical University in Prague, Prague, Czech Republic\\
$^{39}$ Faculty of Science, P.J.~\v{S}af\'{a}rik University, Ko\v{s}ice, Slovakia\\
$^{40}$ Frankfurt Institute for Advanced Studies, Johann Wolfgang Goethe-Universit\"{a}t Frankfurt, Frankfurt, Germany\\
$^{41}$ Fudan University, Shanghai, China\\
$^{42}$ Gangneung-Wonju National University, Gangneung, Republic of Korea\\
$^{43}$ Gauhati University, Department of Physics, Guwahati, India\\
$^{44}$ Helmholtz-Institut f\"{u}r Strahlen- und Kernphysik, Rheinische Friedrich-Wilhelms-Universit\"{a}t Bonn, Bonn, Germany\\
$^{45}$ Helsinki Institute of Physics (HIP), Helsinki, Finland\\
$^{46}$ High Energy Physics Group,  Universidad Aut\'{o}noma de Puebla, Puebla, Mexico\\
$^{47}$ Hiroshima University, Hiroshima, Japan\\
$^{48}$ Hochschule Worms, Zentrum  f\"{u}r Technologietransfer und Telekommunikation (ZTT), Worms, Germany\\
$^{49}$ Horia Hulubei National Institute of Physics and Nuclear Engineering, Bucharest, Romania\\
$^{50}$ Indian Institute of Technology Bombay (IIT), Mumbai, India\\
$^{51}$ Indian Institute of Technology Indore, Indore, India\\
$^{52}$ Indonesian Institute of Sciences, Jakarta, Indonesia\\
$^{53}$ INFN, Laboratori Nazionali di Frascati, Frascati, Italy\\
$^{54}$ INFN, Sezione di Bari, Bari, Italy\\
$^{55}$ INFN, Sezione di Bologna, Bologna, Italy\\
$^{56}$ INFN, Sezione di Cagliari, Cagliari, Italy\\
$^{57}$ INFN, Sezione di Catania, Catania, Italy\\
$^{58}$ INFN, Sezione di Padova, Padova, Italy\\
$^{59}$ INFN, Sezione di Pavia, Pavia, Italy\\
$^{60}$ INFN, Sezione di Roma, Rome, Italy\\
$^{61}$ INFN, Sezione di Torino, Turin, Italy\\
$^{62}$ INFN, Sezione di Trieste, Trieste, Italy\\
$^{63}$ Inha University, Incheon, Republic of Korea\\
$^{64}$ Institute for Gravitational and Subatomic Physics (GRASP), Utrecht University/Nikhef, Utrecht, Netherlands\\
$^{65}$ Institute for Nuclear Research, Academy of Sciences, Moscow, Russia\\
$^{66}$ Institute of Experimental Physics, Slovak Academy of Sciences, Ko\v{s}ice, Slovakia\\
$^{67}$ Institute of Physics, Homi Bhabha National Institute, Bhubaneswar, India\\
$^{68}$ Institute of Physics of the Czech Academy of Sciences, Prague, Czech Republic\\
$^{69}$ Institute of Space Science (ISS), Bucharest, Romania\\
$^{70}$ Institut f\"{u}r Kernphysik, Johann Wolfgang Goethe-Universit\"{a}t Frankfurt, Frankfurt, Germany\\
$^{71}$ Instituto de Ciencias Nucleares, Universidad Nacional Aut\'{o}noma de M\'{e}xico, Mexico City, Mexico\\
$^{72}$ Instituto de F\'{i}sica, Universidade Federal do Rio Grande do Sul (UFRGS), Porto Alegre, Brazil\\
$^{73}$ Instituto de F\'{\i}sica, Universidad Nacional Aut\'{o}noma de M\'{e}xico, Mexico City, Mexico\\
$^{74}$ iThemba LABS, National Research Foundation, Somerset West, South Africa\\
$^{75}$ Jeonbuk National University, Jeonju, Republic of Korea\\
$^{76}$ Johann-Wolfgang-Goethe Universit\"{a}t Frankfurt Institut f\"{u}r Informatik, Fachbereich Informatik und Mathematik, Frankfurt, Germany\\
$^{77}$ Joint Institute for Nuclear Research (JINR), Dubna, Russia\\
$^{78}$ Korea Institute of Science and Technology Information, Daejeon, Republic of Korea\\
$^{79}$ KTO Karatay University, Konya, Turkey\\
$^{80}$ Laboratoire de Physique des 2 Infinis, Ir\`{e}ne Joliot-Curie, Orsay, France\\
$^{81}$ Laboratoire de Physique Subatomique et de Cosmologie, Universit\'{e} Grenoble-Alpes, CNRS-IN2P3, Grenoble, France\\
$^{82}$ Lawrence Berkeley National Laboratory, Berkeley, California, United States\\
$^{83}$ Lund University Department of Physics, Division of Particle Physics, Lund, Sweden\\
$^{84}$ Moscow Institute for Physics and Technology, Moscow, Russia\\
$^{85}$ Nagasaki Institute of Applied Science, Nagasaki, Japan\\
$^{86}$ Nara Women{'}s University (NWU), Nara, Japan\\
$^{87}$ National and Kapodistrian University of Athens, School of Science, Department of Physics , Athens, Greece\\
$^{88}$ National Centre for Nuclear Research, Warsaw, Poland\\
$^{89}$ National Institute of Science Education and Research, Homi Bhabha National Institute, Jatni, India\\
$^{90}$ National Nuclear Research Center, Baku, Azerbaijan\\
$^{91}$ National Research Centre Kurchatov Institute, Moscow, Russia\\
$^{92}$ Niels Bohr Institute, University of Copenhagen, Copenhagen, Denmark\\
$^{93}$ Nikhef, National institute for subatomic physics, Amsterdam, Netherlands\\
$^{94}$ NRC Kurchatov Institute IHEP, Protvino, Russia\\
$^{95}$ NRC \guillemotleft Kurchatov\guillemotright  Institute - ITEP, Moscow, Russia\\
$^{96}$ NRNU Moscow Engineering Physics Institute, Moscow, Russia\\
$^{97}$ Nuclear Physics Group, STFC Daresbury Laboratory, Daresbury, United Kingdom\\
$^{98}$ Nuclear Physics Institute of the Czech Academy of Sciences, \v{R}e\v{z} u Prahy, Czech Republic\\
$^{99}$ Oak Ridge National Laboratory, Oak Ridge, Tennessee, United States\\
$^{100}$ Ohio State University, Columbus, Ohio, United States\\
$^{101}$ Petersburg Nuclear Physics Institute, Gatchina, Russia\\
$^{102}$ Physics department, Faculty of science, University of Zagreb, Zagreb, Croatia\\
$^{103}$ Physics Department, Panjab University, Chandigarh, India\\
$^{104}$ Physics Department, University of Jammu, Jammu, India\\
$^{105}$ Physics Department, University of Rajasthan, Jaipur, India\\
$^{106}$ Physikalisches Institut, Eberhard-Karls-Universit\"{a}t T\"{u}bingen, T\"{u}bingen, Germany\\
$^{107}$ Physikalisches Institut, Ruprecht-Karls-Universit\"{a}t Heidelberg, Heidelberg, Germany\\
$^{108}$ Physik Department, Technische Universit\"{a}t M\"{u}nchen, Munich, Germany\\
$^{109}$ Politecnico di Bari and Sezione INFN, Bari, Italy\\
$^{110}$ Research Division and ExtreMe Matter Institute EMMI, GSI Helmholtzzentrum f\"ur Schwerionenforschung GmbH, Darmstadt, Germany\\
$^{111}$ Russian Federal Nuclear Center (VNIIEF), Sarov, Russia\\
$^{112}$ Saha Institute of Nuclear Physics, Homi Bhabha National Institute, Kolkata, India\\
$^{113}$ School of Physics and Astronomy, University of Birmingham, Birmingham, United Kingdom\\
$^{114}$ Secci\'{o}n F\'{\i}sica, Departamento de Ciencias, Pontificia Universidad Cat\'{o}lica del Per\'{u}, Lima, Peru\\
$^{115}$ St. Petersburg State University, St. Petersburg, Russia\\
$^{116}$ Stefan Meyer Institut f\"{u}r Subatomare Physik (SMI), Vienna, Austria\\
$^{117}$ SUBATECH, IMT Atlantique, Universit\'{e} de Nantes, CNRS-IN2P3, Nantes, France\\
$^{118}$ Suranaree University of Technology, Nakhon Ratchasima, Thailand\\
$^{119}$ Technical University of Ko\v{s}ice, Ko\v{s}ice, Slovakia\\
$^{120}$ The Henryk Niewodniczanski Institute of Nuclear Physics, Polish Academy of Sciences, Cracow, Poland\\
$^{121}$ The University of Texas at Austin, Austin, Texas, United States\\
$^{122}$ Universidad Aut\'{o}noma de Sinaloa, Culiac\'{a}n, Mexico\\
$^{123}$ Universidade de S\~{a}o Paulo (USP), S\~{a}o Paulo, Brazil\\
$^{124}$ Universidade Estadual de Campinas (UNICAMP), Campinas, Brazil\\
$^{125}$ Universidade Federal do ABC, Santo Andre, Brazil\\
$^{126}$ University of Cape Town, Cape Town, South Africa\\
$^{127}$ University of Houston, Houston, Texas, United States\\
$^{128}$ University of Jyv\"{a}skyl\"{a}, Jyv\"{a}skyl\"{a}, Finland\\
$^{129}$ University of Kansas, Lawrence, Kansas, United States\\
$^{130}$ University of Liverpool, Liverpool, United Kingdom\\
$^{131}$ University of Science and Technology of China, Hefei, China\\
$^{132}$ University of South-Eastern Norway, Tonsberg, Norway\\
$^{133}$ University of Tennessee, Knoxville, Tennessee, United States\\
$^{134}$ University of the Witwatersrand, Johannesburg, South Africa\\
$^{135}$ University of Tokyo, Tokyo, Japan\\
$^{136}$ University of Tsukuba, Tsukuba, Japan\\
$^{137}$ Universit\'{e} Clermont Auvergne, CNRS/IN2P3, LPC, Clermont-Ferrand, France\\
$^{138}$ Universit\'{e} de Lyon, CNRS/IN2P3, Institut de Physique des 2 Infinis de Lyon , Lyon, France\\
$^{139}$ Universit\'{e} de Strasbourg, CNRS, IPHC UMR 7178, F-67000 Strasbourg, France, Strasbourg, France\\
$^{140}$ Universit\'{e} Paris-Saclay Centre d'Etudes de Saclay (CEA), IRFU, D\'{e}partment de Physique Nucl\'{e}aire (DPhN), Saclay, France\\
$^{141}$ Universit\`{a} degli Studi di Foggia, Foggia, Italy\\
$^{142}$ Universit\`{a} di Brescia, Brescia, Italy\\
$^{143}$ Variable Energy Cyclotron Centre, Homi Bhabha National Institute, Kolkata, India\\
$^{144}$ Warsaw University of Technology, Warsaw, Poland\\
$^{145}$ Wayne State University, Detroit, Michigan, United States\\
$^{146}$ Westf\"{a}lische Wilhelms-Universit\"{a}t M\"{u}nster, Institut f\"{u}r Kernphysik, M\"{u}nster, Germany\\
$^{147}$ Wigner Research Centre for Physics, Budapest, Hungary\\
$^{148}$ Yale University, New Haven, Connecticut, United States\\
$^{149}$ Yonsei University, Seoul, Republic of Korea\\

\end{flushleft} 

%% file: main.bbl
\providecommand{\href}[2]{#2}\begingroup\raggedright\begin{thebibliography}{10}

\bibitem{Kogler_2019}
R.~Kogler {\em et~al.}, ``{Jet substructure at the Large Hadron Collider}'',
  \href{http://dx.doi.org/10.1103/revmodphys.91.045003}{{\em Rev. Mod. Phys.}
  {\bfseries 91} (Dec, 2019) 045003},
  \href{http://arxiv.org/abs/1803.06991}{{\ttfamily arXiv:1803.06991
  [hep-ex]}}.

\bibitem{Larkoski_2020}
A.~J. Larkoski, I.~Moult, and B.~Nachman, ``{Jet substructure at the Large
  Hadron Collider: A review of recent advances in theory and machine
  learning}'', \href{http://dx.doi.org/10.1016/j.physrep.2019.11.001}{{\em
  Phys. Rept.} {\bfseries 841} (Jan, 2020) 1–63},
  \href{http://arxiv.org/abs/1709.04464}{{\ttfamily arXiv:1709.04464
  [hep-ph]}}.

\bibitem{Larkoski:2014wba}
A.~J. Larkoski, S.~Marzani, G.~Soyez, and J.~Thaler, ``{Soft Drop}'',
  \href{http://dx.doi.org/10.1007/JHEP05(2014)146}{{\em JHEP} {\bfseries 05}
  (May, 2014) 146}, \href{http://arxiv.org/abs/1402.2657}{{\ttfamily
  arXiv:1402.2657 [hep-ph]}}.

\bibitem{Dasgupta:2013ihk}
M.~Dasgupta, A.~Fregoso, S.~Marzani, and G.~P. Salam, ``{Towards an
  understanding of jet substructure}'',
  \href{http://dx.doi.org/10.1007/JHEP09(2013)029}{{\em JHEP} {\bfseries 09}
  (Sep, 2013) 029}, \href{http://arxiv.org/abs/1307.0007}{{\ttfamily
  arXiv:1307.0007 [hep-ph]}}.

\bibitem{Larkoski:2015lea}
A.~J. Larkoski, S.~Marzani, and J.~Thaler, ``{Sudakov Safety in Perturbative
  QCD}'', \href{http://dx.doi.org/10.1103/PhysRevD.91.111501}{{\em Phys. Rev.
  D} {\bfseries 91} (Jun, 2015) 111501},
  \href{http://arxiv.org/abs/1502.01719}{{\ttfamily arXiv:1502.01719
  [hep-ph]}}.

\bibitem{Larkoski_2014}
A.~J. Larkoski, J.~Thaler, and W.~J. Waalewijn, ``Gaining (mutual) information
  about quark/gluon discrimination'',
  \href{http://dx.doi.org/10.1007/jhep11(2014)129}{{\em JHEP} {\bfseries 11}
  (Nov, 2014) 129}, \href{http://arxiv.org/abs/1408.3122}{{\ttfamily
  arXiv:1408.3122 [hep-ph]}}.

\bibitem{Almeida:2008yp}
L.~G. Almeida, S.~J. Lee, G.~Perez, G.~F. Sterman, I.~Sung, and J.~Virzi,
  ``{Substructure of high-$p_\text{T}$ Jets at the LHC}'',
  \href{http://dx.doi.org/10.1103/PhysRevD.79.074017}{{\em Phys. Rev. D}
  {\bfseries 79} (Apr, 2009) 074017},
  \href{http://arxiv.org/abs/0807.0234}{{\ttfamily arXiv:0807.0234 [hep-ph]}}.

\bibitem{Berger_2003}
C.~F. Berger and G.~Sterman, ``Scaling rule for nonperturbative radiation in a
  class of event shapes'',
  \href{http://dx.doi.org/10.1088/1126-6708/2003/09/058}{{\em JHEP} {\bfseries
  09} (Sep, 2003) 058–058},
  \href{http://arxiv.org/abs/hep-ph/0307394}{{\ttfamily arXiv:hep-ph/0307394}}.

\bibitem{Kang_2018}
Z.-B. Kang, K.~Lee, and F.~Ringer, ``Jet angularity measurements for single
  inclusive jet production'',
  \href{http://dx.doi.org/10.1007/jhep04(2018)110}{{\em JHEP} {\bfseries 04}
  (Apr, 2018) 110}, \href{http://arxiv.org/abs/1801.00790}{{\ttfamily
  arXiv:1801.00790 [hep-ph]}}.

\bibitem{Catani:1992jc}
S.~Catani, G.~Turnock, and B.~R. Webber, ``{Jet broadening measures in $e^{+}
  e^{-}$ annihilation}'',
  \href{http://dx.doi.org/10.1016/0370-2693(92)91565-Q}{{\em Phys. Lett. B}
  {\bfseries 295} (Dec, 1992) 269--276}.

\bibitem{Farhi:1977sg}
E.~Farhi, ``{A QCD Test for Jets}'',
  \href{http://dx.doi.org/10.1103/PhysRevLett.39.1587}{{\em Phys. Rev. Lett.}
  {\bfseries 39} (Sep, 1977) 1587--1588}.

\bibitem{Larkoski:2014tva}
A.~J. Larkoski, I.~Moult, and D.~Neill, ``{Toward Multi-Differential Cross
  Sections: Measuring Two Angularities on a Single Jet}'',
  \href{http://dx.doi.org/10.1007/JHEP09(2014)046}{{\em JHEP} {\bfseries 09}
  (Sep, 2014) 046}, \href{http://arxiv.org/abs/1401.4458}{{\ttfamily
  arXiv:1401.4458 [hep-ph]}}.

\bibitem{Procura:2018zpn}
M.~Procura, W.~J. Waalewijn, and L.~Zeune, ``{Joint resummation of two
  angularities at next-to-next-to-leading logarithmic order}'',
  \href{http://dx.doi.org/10.1007/JHEP10(2018)098}{{\em JHEP} {\bfseries 10}
  (Oct, 2018) 098}, \href{http://arxiv.org/abs/1806.10622}{{\ttfamily
  arXiv:1806.10622 [hep-ph]}}.

\bibitem{Stewart:2014nna}
I.~W. Stewart, F.~J. Tackmann, and W.~J. Waalewijn, ``{Dissecting Soft
  Radiation with Factorization}'',
  \href{http://dx.doi.org/10.1103/PhysRevLett.114.092001}{{\em Phys. Rev.
  Lett.} {\bfseries 114} (Mar, 2015) 092001},
  \href{http://arxiv.org/abs/1405.6722}{{\ttfamily arXiv:1405.6722 [hep-ph]}}.

\bibitem{Lee_2007}
C.~Lee and G.~Sterman, ``{Momentum flow correlations from event shapes:
  Factorized soft gluons and soft-collinear effective theory}'',
  \href{http://dx.doi.org/10.1103/physrevd.75.014022}{{\em Phys. Rev. D}
  {\bfseries 75} (Jan, 2007) },
  \href{http://arxiv.org/abs/hep-ph/0611061}{{\ttfamily arXiv:hep-ph/0611061}}.

\bibitem{Chang:2013rca}
H.-M. Chang, M.~Procura, J.~Thaler, and W.~J. Waalewijn, ``{Calculating
  Track-Based Observables for the LHC}'',
  \href{http://dx.doi.org/10.1103/PhysRevLett.111.102002}{{\em Phys. Rev.
  Lett.} {\bfseries 111} (Sep, 2013) 102002},
  \href{http://arxiv.org/abs/1303.6637}{{\ttfamily arXiv:1303.6637 [hep-ph]}}.

\bibitem{KANG201941}
Z.-B. Kang, K.~Lee, X.~Liu, and F.~Ringer, ``{Soft drop groomed jet
  angularities at the LHC}'',
  \href{http://dx.doi.org/10.1016/j.physletb.2019.04.018}{{\em Phys. Lett. B}
  {\bfseries 793} (Jun, 2019) 41 -- 47},
  \href{http://arxiv.org/abs/1811.06983}{{\ttfamily arXiv:1811.06983
  [hep-ph]}}.

\bibitem{Caletti:2021oor}
S.~Caletti, O.~Fedkevych, S.~Marzani, D.~Reichelt, S.~Schumann, G.~Soyez, and
  V.~Theeuwes, ``{Jet angularities in Z+jet production at the LHC}'',
  \href{http://dx.doi.org/10.1007/JHEP07(2021)076}{{\em JHEP} {\bfseries 07}
  (2021) 076}, \href{http://arxiv.org/abs/2104.06920}{{\ttfamily
  arXiv:2104.06920 [hep-ph]}}.

\bibitem{Almeida:2014uva}
L.~G. Almeida, S.~D. Ellis, C.~Lee, G.~Sterman, I.~Sung, and J.~R. Walsh,
  ``{Comparing and counting logs in direct and effective methods of QCD
  resummation}'', \href{http://dx.doi.org/10.1007/JHEP04(2014)174}{{\em JHEP}
  {\bfseries 04} (Apr, 2014) 174},
  \href{http://arxiv.org/abs/1401.4460}{{\ttfamily arXiv:1401.4460 [hep-ph]}}.

\bibitem{Korchemsky_1999_Fnp}
G.~P. Korchemsky and G.~Sterman, ``Power corrections to event shapes and
  factorization'', \href{http://dx.doi.org/10.1016/s0550-3213(99)00308-9}{{\em
  Nuclear Physics B} {\bfseries 555} (Aug, 1999) 335–351},
  \href{http://arxiv.org/abs/hep-ph/9902341}{{\ttfamily arXiv:hep-ph/9902341}}.

\bibitem{Aschenauer_2020_Fnp}
E.-C. Aschenauer, K.~Lee, B.~Page, and F.~Ringer, ``{Jet angularities in
  photoproduction at the Electron-Ion Collider}'',
  \href{http://dx.doi.org/10.1103/physrevd.101.054028}{{\em Phys. Rev. D}
  {\bfseries 101} (Mar, 2020) },
  \href{http://arxiv.org/abs/1910.11460}{{\ttfamily arXiv:1910.11460
  [hep-ph]}}.

\bibitem{ang2018}
{\bfseries {ALICE}} Collaboration, ``{Medium modification of the shape of
  small-radius jets in central Pb-Pb collisions at
  $\sqrt{s_{\mathrm{NN}}}=2.76$ TeV}'',
  \href{http://dx.doi.org/10.1007/jhep10(2018)139}{{\em JHEP} {\bfseries 10}
  (Oct, 2018) 139}, \href{http://arxiv.org/abs/1807.06854}{{\ttfamily
  arXiv:1807.06854 [nucl-ex]}}.

\bibitem{Aad_2012}
{\bfseries ATLAS} Collaboration, G.~Aad {\em et~al.}, ``{ATLAS measurements of
  the properties of jets for boosted particle searches}'',
  \href{http://dx.doi.org/10.1103/physrevd.86.072006}{{\em Phys. Rev. D}
  {\bfseries 86} (Oct, 2012) 072006},
  \href{http://arxiv.org/abs/1206.5369}{{\ttfamily arXiv:1206.5369 [hep-ex]}}.

\bibitem{PhysRevD.98.092014}
{\bfseries CMS} Collaboration, A.~M. Sirunyan {\em et~al.}, ``{Measurement of
  jet substructure observables in $t\overline{t}$ events from proton-proton
  collisions at $\sqrt{s}=13$ TeV}'',
  \href{http://dx.doi.org/10.1103/PhysRevD.98.092014}{{\em Phys. Rev. D}
  {\bfseries 98} (Nov, 2018) 092014},
  \href{http://arxiv.org/abs/1808.07340}{{\ttfamily arXiv:1808.07340
  [hep-ex]}}.

\bibitem{Aaltonen:2011pg}
{\bfseries CDF} Collaboration, T.~Aaltonen {\em et~al.}, ``{Study of
  Substructure of High Transverse Momentum Jets Produced in Proton-Antiproton
  Collisions at $\sqrt{s}=1.96$ TeV}'',
  \href{http://dx.doi.org/10.1103/PhysRevD.85.091101}{{\em Phys. Rev. D}
  {\bfseries 85} (May, 2012) 091101},
  \href{http://arxiv.org/abs/1106.5952}{{\ttfamily arXiv:1106.5952 [hep-ex]}}.

\bibitem{ATLAS:2012am}
{\bfseries ATLAS} Collaboration, G.~Aad {\em et~al.}, ``{Jet mass and
  substructure of inclusive jets in $\sqrt{s}=7$ TeV $pp$ collisions with the
  ATLAS experiment}'', \href{http://dx.doi.org/10.1007/JHEP05(2012)128}{{\em
  JHEP} {\bfseries 05} (May, 2012) 128},
  \href{http://arxiv.org/abs/1203.4606}{{\ttfamily arXiv:1203.4606 [hep-ex]}}.

\bibitem{Aad:2019wdr}
{\bfseries ATLAS} Collaboration, G.~Aad {\em et~al.}, ``{Measurement of the jet
  mass in high transverse momentum $Z(\rightarrow b\overline{b})\gamma$
  production at $\sqrt{s}= 13$ TeV using the ATLAS detector}'',
  \href{http://dx.doi.org/10.1016/j.physletb.2020.135991}{{\em Phys. Lett. B}
  {\bfseries 812} (Jan, 2021) 135991},
  \href{http://arxiv.org/abs/1907.07093}{{\ttfamily arXiv:1907.07093
  [hep-ex]}}.

\bibitem{PhysRevD.101.052007}
{\bfseries ATLAS} Collaboration, G.~Aad {\em et~al.}, ``{Measurement of
  soft-drop jet observables in \pp{} collisions with the {ATLAS} detector at
  $\sqrts=13$ TeV}'', \href{http://dx.doi.org/10.1103/PhysRevD.101.052007}{{\em
  Phys. Rev. D} {\bfseries 101} (Mar, 2020) 052007},
  \href{http://arxiv.org/abs/1912.09837}{{\ttfamily arXiv:1912.09837
  [hep-ex]}}.

\bibitem{Sirunyan:2019rfa}
{\bfseries CMS} Collaboration, A.~M. Sirunyan {\em et~al.}, ``{Measurement of
  the Jet Mass Distribution and Top Quark Mass in Hadronic Decays of Boosted
  Top Quarks in \pp{} Collisions at $\sqrt{s} = 13$ TeV}'',
  \href{http://dx.doi.org/10.1103/PhysRevLett.124.202001}{{\em Phys. Rev.
  Lett.} {\bfseries 124} (May, 2020) 202001},
  \href{http://arxiv.org/abs/1911.03800}{{\ttfamily arXiv:1911.03800
  [hep-ex]}}.

\bibitem{Sirunyan2018}
{\bfseries CMS} Collaboration, A.~M. Sirunyan {\em et~al.}, ``{Measurement of
  the groomed jet mass in \PbPb{} and \pp{} collisions at $\sqrts=5.02$ TeV}'',
  \href{http://dx.doi.org/10.1007/JHEP10(2018)161}{{\em JHEP} {\bfseries 10}
  (Oct, 2018) 161}, \href{http://arxiv.org/abs/1805.05145}{{\ttfamily
  arXiv:1805.05145 [hep-ex]}}.

\bibitem{Sirunyan_2017}
{\bfseries CMS} Collaboration, A.~M. Sirunyan {\em et~al.}, ``Measurement of
  the jet mass in highly boosted ${\mathrm{t}}\overline{\mathrm{t}}$ events
  from pp collisions at $\sqrt{s}=8\,\text{TeV}$'',
  \href{http://dx.doi.org/10.1140/epjc/s10052-017-5030-3}{{\em Eur. Phys. J. C}
  {\bfseries 77} (Jul, 2017) 467},
  \href{http://arxiv.org/abs/1703.06330}{{\ttfamily arXiv:1703.06330
  [hep-ex]}}.

\bibitem{CMS_dijet_mass_2013}
{\bfseries CMS} Collaboration, S.~Chatrchyan {\em et~al.}, ``{Studies of jet
  mass in dijet and W/Z + jet events}'',
  \href{http://dx.doi.org/10.1007/jhep05(2013)090}{{\em JHEP} {\bfseries 05}
  (May, 2013) 090}, \href{http://arxiv.org/abs/1303.4811}{{\ttfamily
  arXiv:1303.4811 [hep-ex]}}.

\bibitem{CMS_pp_mass_2018}
{\bfseries {CMS}} Collaboration, A.~M. Sirunyan {\em et~al.}, ``{Measurements
  of the differential jet cross section as a function of the jet mass in dijet
  events from proton-proton collisions at $\sqrt{s}=13$ TeV}'',
  \href{http://dx.doi.org/10.1007/jhep11(2018)113}{{\em JHEP} {\bfseries 2018}
  (Nov, 2018) 113}, \href{http://arxiv.org/abs/1807.05974}{{\ttfamily
  arXiv:1807.05974 [hep-ex]}}.

\bibitem{ATLAS-CONF-2018-014}
{\bfseries ATLAS} Collaboration, ``{Measurement of $R=0.4$ jet mass in Pb+Pb
  and pp collisions at $\sqrt{s_{\mathrm{NN}}}=5.02$ TeV with the ATLAS
  detector}'', tech. rep., CERN, Geneva, CH, May, 2018.
\newblock \url{https://cds.cern.ch/record/2319867}.

\bibitem{ATLAS_SD_mass_2018}
{\bfseries ATLAS} Collaboration, M.~Aaboud {\em et~al.}, ``{Measurement of the
  soft-drop jet mass in pp collisions at $\sqrt{s}=13$ TeV with the ATLAS
  detector}'', \href{http://dx.doi.org/10.1103/physrevlett.121.092001}{{\em
  Phys. Rev. Lett.} {\bfseries 121} (Aug, 2018) 092001},
  \href{http://arxiv.org/abs/1711.08341}{{\ttfamily arXiv:1711.08341
  [hep-ex]}}.

\bibitem{Acharya:2017goa}
{\bfseries ALICE} Collaboration, S.~Acharya {\em et~al.}, ``{First measurement
  of jet mass in Pb\textendash{}Pb and p\textendash{}Pb collisions at the
  LHC}'', \href{http://dx.doi.org/10.1016/j.physletb.2017.11.044}{{\em Phys.
  Lett. B} {\bfseries 776} (Jan, 2018) 249--264},
  \href{http://arxiv.org/abs/1702.00804}{{\ttfamily arXiv:1702.00804
  [nucl-ex]}}.

\bibitem{Jacak:2012dx}
B.~V. Jacak and B.~M{\"u}ller, ``{The exploration of hot nuclear matter}'',
  \href{http://dx.doi.org/10.1126/science.1215901}{{\em Science} {\bfseries
  337} (Jul, 2012) 310--314}.

\bibitem{LHC1review}
B.~M{\"u}ller, J.~Schukraft, and B.~Wys{\l}ouch, ``{First Results from Pb+Pb
  Collisions at the LHC}'',
  \href{http://dx.doi.org/10.1146/annurev-nucl-102711-094910}{{\em Annu. Rev.
  Nucl. Part. S.} {\bfseries 62} (Nov, 2012) 361--386},
  \href{http://arxiv.org/abs/1202.3233}{{\ttfamily arXiv:1202.3233 [hep-ex]}}.

\bibitem{Braun-Munzinger:2015hba}
P.~Braun-Munzinger, V.~Koch, T.~Schäfer, and J.~Stachel, ``{Properties of hot
  and dense matter from relativistic heavy ion collisions}'',
  \href{http://dx.doi.org/10.1016/j.physrep.2015.12.003}{{\em Phys. Rept.}
  {\bfseries 621} (Mar, 2016) 76--126},
  \href{http://arxiv.org/abs/1510.00442}{{\ttfamily arXiv:1510.00442
  [nucl-th]}}. Memorial Volume in Honor of Gerald E. Brown.

\bibitem{TheBigPicture}
W.~Busza, K.~Rajagopal, and W.~van~der Schee, ``{Heavy Ion Collisions: The Big
  Picture, and the Big Questions}'',
  \href{http://dx.doi.org/10.1146/annurev-nucl-101917-020852}{{\em Annu. Rev.
  Nucl. Part. S.} {\bfseries 68} (Oct, 2018) 339--376},
  \href{http://arxiv.org/abs/1802.04801}{{\ttfamily arXiv:1802.04801
  [hep-ph]}}.

\bibitem{ReviewXinNian}
G.-Y. Qin and X.-N. Wang, ``{Jet quenching in high-energy heavy-ion
  collisions}'', \href{http://dx.doi.org/10.1142/S0218301315300143}{{\em Int.
  J. Mod. Phys. E} {\bfseries 24} (Oct, 2015) 1530014},
  \href{http://arxiv.org/abs/1511.00790}{{\ttfamily arXiv:1511.00790
  [hep-ph]}}.

\bibitem{ReviewYacine}
J.-P. Blaizot and Y.~Mehtar-Tani, ``{Jet structure in heavy ion collisions}'',
  \href{http://dx.doi.org/10.1142/S021830131530012X}{{\em Int. J. Mod. Phys. E}
  {\bfseries 24} (Oct, 2015) 1530012},
  \href{http://arxiv.org/abs/1503.05958}{{\ttfamily arXiv:1503.05958
  [hep-ph]}}.

\bibitem{ReviewMajumder}
A.~Majumder and M.~van Leeuwen, ``{The theory and phenomenology of perturbative
  QCD based jet quenching}'',
  \href{http://dx.doi.org/10.1016/j.ppnp.2010.09.001}{{\em Prog. Part. Nucl.
  Phys.} {\bfseries 66} (Jan, 2011) 41 -- 92},
  \href{http://arxiv.org/abs/1002.2206}{{\ttfamily arXiv:1002.2206 [hep-ph]}}.

\bibitem{Yan:2020zrz}
J.~Yan, S.-Y. Chen, W.~Dai, B.-W. Zhang, and E.~Wang, ``{Medium modifications
  of girth distributions for inclusive jets and $Z^0+{\rm jet}$ in relativistic
  heavy-ion collisions at the LHC}'',
  \href{http://dx.doi.org/10.1088/1674-1137/abca2b}{{\em Chin. Phys. C}
  {\bfseries 45} (Feb, 2021) 024102},
  \href{http://arxiv.org/abs/2005.01093}{{\ttfamily arXiv:2005.01093
  [hep-ph]}}.

\bibitem{Elayavalli2017}
R.~Kunnawalkam~Elayavalli and K.~C. Zapp, ``Medium response in {JEWEL} and its
  impact on jet shape observables in heavy ion collisions'',
  \href{http://dx.doi.org/10.1007/JHEP07(2017)141}{{\em JHEP} {\bfseries 07}
  (Jul, 2017) 141}, \href{http://arxiv.org/abs/1707.01539}{{\ttfamily
  arXiv:1707.01539 [hep-ph]}}.

\bibitem{Casalderrey-Solana:2019ubu}
J.~Casalderrey-Solana, G.~Milhano, D.~Pablos, and K.~Rajagopal, ``{Modification
  of Jet Substructure in Heavy Ion Collisions as a Probe of the Resolution
  Length of Quark-Gluon Plasma}'',
  \href{http://dx.doi.org/10.1007/JHEP01(2020)044}{{\em JHEP} {\bfseries 01}
  (Jan, 2020) 044}, \href{http://arxiv.org/abs/1907.11248}{{\ttfamily
  arXiv:1907.11248 [hep-ph]}}.

\bibitem{Acharya:2019jyg}
{\bfseries ALICE} Collaboration, S.~Acharya {\em et~al.}, ``{Measurements of
  inclusive jet spectra in \pp{} and central \PbPb{} collisions at
  $\sqrt{s_{\rm{NN}}} = 5.02$ TeV}'',
  \href{http://dx.doi.org/10.1103/PhysRevC.101.034911}{{\em Phys. Rev. C}
  {\bfseries 101} (Mar, 2020) 034911},
  \href{http://arxiv.org/abs/1909.09718}{{\ttfamily arXiv:1909.09718
  [nucl-ex]}}.

\bibitem{Mulligan:2020tim}
J.~Mulligan and M.~P{\l}osko{\'n}, ``{Identifying groomed jet splittings in
  heavy-ion collisions}'',
  \href{http://dx.doi.org/10.1103/PhysRevC.102.044913}{{\em Phys. Rev. C}
  {\bfseries 102} (Oct, 2020) 044913},
  \href{http://arxiv.org/abs/2006.01812}{{\ttfamily arXiv:2006.01812
  [hep-ph]}}.

\bibitem{pythia}
T.~Sj\"ostrand, S.~Ask, J.~R. Christiansen, R.~Corke, N.~Desai, P.~Ilten,
  S.~Mrenna, S.~Prestel, C.~O. Rasmussen, and P.~Z. Skands, ``{An introduction
  to PYTHIA 8.2}'', \href{http://dx.doi.org/10.1016/j.cpc.2015.01.024}{{\em
  Comput. Phys. Commun.} {\bfseries 191} (Jun, 2015) 159 -- 177},
  \href{http://arxiv.org/abs/1410.3012}{{\ttfamily arXiv:1410.3012 [hep-ph]}}.

\bibitem{B_hr_2008}
M.~Bähr, {\em et~al.}, ``{Herwig++ Physics and Manual}'',
  \href{http://dx.doi.org/10.1140/epjc/s10052-008-0798-9}{{\em Eur. Phys. J. C}
  {\bfseries 58} (Nov, 2008) 639–707},
  \href{http://arxiv.org/abs/0803.0883}{{\ttfamily arXiv:0803.0883 [hep-ph]}}.

\bibitem{Bellm:2015jjp}
J.~Bellm {\em et~al.}, ``{Herwig 7.0/Herwig++ 3.0 release note}'',
  \href{http://dx.doi.org/10.1140/epjc/s10052-016-4018-8}{{\em Eur. Phys. J. C}
  {\bfseries 76} (Apr, 2016) 196},
  \href{http://arxiv.org/abs/1512.01178}{{\ttfamily arXiv:1512.01178
  [hep-ph]}}.

\bibitem{aliceDetector}
{\bfseries ALICE} Collaboration, K.~Aamodt {\em et~al.}, ``{The ALICE
  experiment at the CERN LHC}'',
  \href{http://dx.doi.org/10.1088/1748-0221/3/08/S08002}{{\em JINST} {\bfseries
  3} (Aug, 2008) }.

\bibitem{alicePerformance}
{\bfseries ALICE} Collaboration, B.~Abelev {\em et~al.}, ``{Performance of the
  ALICE Experiment at the CERN LHC}'',
  \href{http://dx.doi.org/10.1142/S0217751X14300440}{{\em Int. J. Mod. Phys.}
  {\bfseries A29} (Sep, 2014) 1430044},
  \href{http://arxiv.org/abs/1402.4476}{{\ttfamily arXiv:1402.4476 [nucl-ex]}}.

\bibitem{LHCmachine}
L.~Evans and P.~Bryant, ``{The CERN Large Hadron Collider: accelerator and
  experiments}'', \href{http://dx.doi.org/10.1088/1748-0221/3/08/s08001}{{\em
  JINST} {\bfseries 3} (Aug, 2008) S08001}.

\bibitem{Abbas:2013taa}
{\bfseries ALICE} Collaboration, E.~Abbas {\em et~al.}, ``{Performance of the
  ALICE VZERO system}'',
  \href{http://dx.doi.org/10.1088/1748-0221/8/10/P10016}{{\em JINST} {\bfseries
  8} (Oct, 2013) P10016}, \href{http://arxiv.org/abs/1306.3130}{{\ttfamily
  arXiv:1306.3130 [nucl-ex]}}.

\bibitem{ppXsec}
{\bfseries ALICE} Collaboration, S.~Acharya {\em et~al.}, ``{ALICE 2017
  luminosity determination for pp collisions at $\sqrt{s} = 5$ TeV}'',
  ALICE-PUBLIC-2018-014. \url{https://cds.cern.ch/record/2648933}.

\bibitem{Alme_2010}
{\bfseries ALICE} Collaboration, J.~Alme {\em et~al.}, ``{The ALICE TPC, a
  large 3-dimensional tracking device with fast readout for ultra-high
  multiplicity events}'',
  \href{http://dx.doi.org/10.1016/j.nima.2010.04.042}{{\em Nuclear Instruments
  and Methods in Physics Research Section A: Accelerators, Spectrometers,
  Detectors and Associated Equipment} {\bfseries 622} (Oct, 2010) 316--367},
  \href{http://arxiv.org/abs/1001.1950}{{\ttfamily arXiv:1001.1950
  [physics.ins-det]}}.

\bibitem{Aamodt:2010aa}
{\bfseries ALICE} Collaboration, K.~Aamodt {\em et~al.}, ``{Alignment of the
  ALICE Inner Tracking System with cosmic-ray tracks}'',
  \href{http://dx.doi.org/10.1088/1748-0221/5/03/P03003}{{\em JINST} {\bfseries
  5} (Mar, 2010) P03003}, \href{http://arxiv.org/abs/1001.0502}{{\ttfamily
  arXiv:1001.0502 [physics.ins-det]}}.

\bibitem{Acharya:2019tku}
{\bfseries ALICE} Collaboration, S.~Acharya {\em et~al.}, ``{Measurement of
  charged jet cross section in $pp$ collisions at ${\sqrt{s}=5.02}$ TeV}'',
  \href{http://dx.doi.org/10.1103/PhysRevD.100.092004}{{\em Phys. Rev. D}
  {\bfseries 100} (Dec, 2019) 092004},
  \href{http://arxiv.org/abs/1905.02536}{{\ttfamily arXiv:1905.02536
  [nucl-ex]}}.

\bibitem{Brun:1119728}
R.~Brun, F.~Bruyant, M.~Maire, A.~C. McPherson, and P.~Zanarini, {\em {GEANT 3:
  user's guide Geant 3.10, Geant 3.11; rev. version}}.
\newblock CERN, Geneva, Sep, 1987.
\newblock \url{https://cds.cern.ch/record/1119728}.

\bibitem{Cacciari:2011ma}
M.~Cacciari, G.~P. Salam, and G.~Soyez, ``{FastJet User Manual}'',
  \href{http://dx.doi.org/10.1140/epjc/s10052-012-1896-2}{{\em Eur. Phys. J. C}
  {\bfseries 72} (Mar, 2012) 1896},
  \href{http://arxiv.org/abs/1111.6097}{{\ttfamily arXiv:1111.6097 [hep-ph]}}.

\bibitem{antikt}
M.~Cacciari, G.~P. Salam, and G.~Soyez, ``{The anti-$k_t$ jet clustering
  algorithm}'', \href{http://dx.doi.org/10.1088/1126-6708/2008/04/063}{{\em
  JHEP} {\bfseries 04} (Apr, 2008) 063},
  \href{http://arxiv.org/abs/0802.1189}{{\ttfamily arXiv:0802.1189 [hep-ph]}}.

\bibitem{Abelev:2013kqa}
{\bfseries ALICE} Collaboration, B.~Abelev {\em et~al.}, ``{Measurement of
  charged jet suppression in Pb-Pb collisions at $\sqrt{s_{NN}} = 2.76$ TeV}'',
  \href{http://dx.doi.org/10.1007/JHEP03(2014)013}{{\em JHEP} {\bfseries 03}
  (Mar, 2014) 013}, \href{http://arxiv.org/abs/1311.0633}{{\ttfamily
  arXiv:1311.0633 [nucl-ex]}}.

\bibitem{Dokshitzer_1997}
Y.~Dokshitzer, G.~Leder, S.~Moretti, and B.~Webber, ``Better jet clustering
  algorithms'', \href{http://dx.doi.org/10.1088/1126-6708/1997/08/001}{{\em
  JHEP} {\bfseries 08} (Aug, 1997) 001}.

\bibitem{Cal_2020}
P.~Cal, K.~Lee, F.~Ringer, and W.~J. Waalewijn, ``Jet energy drop'',
  \href{http://dx.doi.org/10.1007/jhep11(2020)012}{{\em JHEP} {\bfseries 2020}
  (Nov, 2020) }, \href{http://arxiv.org/abs/2007.12187}{{\ttfamily
  arXiv:2007.12187 [hep-ph]}}.

\bibitem{Skands_2014}
P.~Skands, S.~Carrazza, and J.~Rojo, ``{Tuning PYTHIA 8.1: the Monash 2013
  tune}'', \href{http://dx.doi.org/10.1140/epjc/s10052-014-3024-y}{{\em Eur.
  Phys. J. C} {\bfseries 74} (Aug, 2014) },
  \href{http://arxiv.org/abs/1404.5630}{{\ttfamily arXiv:1404.5630 [hep-ph]}}.

\bibitem{primaryParticleALICE}
{\bfseries ALICE} Collaboration, S.~Acharya {\em et~al.}, ``{The ALICE
  definition of primary particles}'', ALICE-PUBLIC-2017-005.
  \url{https://cds.cern.ch/record/2270008}.

\bibitem{DAgostini}
G.~D'Agostini, ``{A multidimensional unfolding method based on Bayes'
  theorem}'', \href{http://dx.doi.org/10.1016/0168-9002(95)00274-X}{{\em
  Nuclear Instruments and Methods in Physics Research Section A: Accelerators,
  Spectrometers, Detectors and Associated Equipment} {\bfseries 362} (Jun,
  1994) 487 -- 498}.

\bibitem{dagostini2010improved}
G.~D'Agostini, ``{Improved iterative Bayesian unfolding}'',
  \href{http://arxiv.org/abs/1010.0632}{{\ttfamily arXiv:1010.0632
  [physics.data-an]}}.

\bibitem{roounfold}
``{RooUnfold}.''
\newblock \url{https://hepunx.rl.ac.uk/~adye/software/unfold/RooUnfold.html}.

\bibitem{protonKpi_ALICE}
{\bfseries ALICE} Collaboration, S.~Acharya {\em et~al.}, ``{Production of
  charged pions, kaons, and (anti-)protons in Pb-Pb and inelastic pp collisions
  at $\sqrt{s_\text{NN}}=5.02$ TeV}'',
  \href{http://dx.doi.org/10.1103/physrevc.101.044907}{{\em Phys. Rev. C}
  {\bfseries 101} (Apr, 2020) 044907}.

\bibitem{Dasgupta:2001sh}
M.~Dasgupta and G.~P. Salam, ``{Resummation of nonglobal QCD observables}'',
  \href{http://dx.doi.org/10.1016/S0370-2693(01)00725-0}{{\em Phys. Lett. B}
  {\bfseries 512} (Jul, 2001) 323--330},
  \href{http://arxiv.org/abs/hep-ph/0104277}{{\ttfamily arXiv:hep-ph/0104277}}.

\bibitem{SCETplaceholder}
C.~W. Bauer, D.~Pirjol, and I.~W. Stewart, ``{Soft collinear factorization in
  effective field theory}'',
  \href{http://dx.doi.org/10.1103/PhysRevD.65.054022}{{\em Phys. Rev. D}
  {\bfseries 65} (Feb, 2002) 054022},
  \href{http://arxiv.org/abs/hep-ph/0109045}{{\ttfamily arXiv:hep-ph/0109045}}.

\bibitem{Chen:2020vvp}
H.~Chen, I.~Moult, X.~Zhang, and H.~X. Zhu, ``{Rethinking jets with energy
  correlators: Tracks, resummation, and analytic continuation}'',
  \href{http://dx.doi.org/10.1103/PhysRevD.102.054012}{{\em Phys. Rev. D}
  {\bfseries 102} (Sep, 2020) 054012},
  \href{http://arxiv.org/abs/2004.11381}{{\ttfamily arXiv:2004.11381
  [hep-ph]}}.

\bibitem{Chien:2020hzh}
Y.-T. Chien, R.~Rahn, S.~Schrijnder~van Velzen, D.~Y. Shao, W.~J. Waalewijn,
  and B.~Wu, ``{Recoil-free azimuthal angle for precision boson-jet
  correlation}'',
  \href{http://dx.doi.org/https://doi.org/10.1016/j.physletb.2021.136124}{{\em
  Phys. Lett. B} {\bfseries 815} (Apr, 2021) 136124},
  \href{http://arxiv.org/abs/2005.12279}{{\ttfamily arXiv:2005.12279
  [hep-ph]}}.

\bibitem{Gieseke:2012ft}
S.~Gieseke, C.~Rohr, and A.~Siodmok, ``{Colour reconnections in Herwig++}'',
  \href{http://dx.doi.org/10.1140/epjc/s10052-012-2225-5}{{\em Eur. Phys. J. C}
  {\bfseries 72} (Nov, 2012) 2225},
  \href{http://arxiv.org/abs/1206.0041}{{\ttfamily arXiv:1206.0041 [hep-ph]}}.

\bibitem{Kang_2016}
Z.-B. Kang, F.~Ringer, and I.~Vitev, ``{The semi-inclusive jet function in SCET
  and small radius resummation for inclusive jet production}'',
  \href{http://dx.doi.org/10.1007/jhep10(2016)125}{{\em JHEP} {\bfseries 10}
  (Oct, 2016) 125}, \href{http://arxiv.org/abs/1606.06732}{{\ttfamily
  arXiv:1606.06732 [hep-ph]}}.

\bibitem{Hoang:2019ceu}
A.~H. Hoang, S.~Mantry, A.~Pathak, and I.~W. Stewart, ``{Nonperturbative
  Corrections to Soft Drop Jet Mass}'',
  \href{http://dx.doi.org/10.1007/JHEP12(2019)002}{{\em JHEP} {\bfseries 12}
  (Dec, 2019) 002}, \href{http://arxiv.org/abs/1906.11843}{{\ttfamily
  arXiv:1906.11843 [hep-ph]}}.

\bibitem{Pathak:2020iue}
A.~Pathak, I.~W. Stewart, V.~Vaidya, and L.~Zoppi, ``{EFT for Soft Drop Double
  Differential Cross Section}'',
  \href{http://dx.doi.org/10.1007/JHEP04(2021)032}{{\em JHEP} {\bfseries 04}
  (Apr, 2021) 32}, \href{http://arxiv.org/abs/2012.15568}{{\ttfamily
  arXiv:2012.15568 [hep-ph]}}.

\bibitem{Kang:2018jwa}
Z.-B. Kang, K.~Lee, X.~Liu, and F.~Ringer, ``{The groomed and ungroomed jet
  mass distribution for inclusive jet production at the LHC}'',
  \href{http://dx.doi.org/10.1007/JHEP10(2018)137}{{\em JHEP} {\bfseries 10}
  (Oct, 2018) 137}, \href{http://arxiv.org/abs/1803.03645}{{\ttfamily
  arXiv:1803.03645 [hep-ph]}}.

\end{thebibliography}\endgroup
